\documentclass[aps,prb,superscriptaddress,citeautoscript, twocolumn]{revtex4}
\usepackage{graphics}
\usepackage{graphicx}
\usepackage{amsmath}
\usepackage{amsfonts}
\usepackage{epsfig}
\usepackage{comment,braket}
\usepackage{color}

\def\beq{\begin{equation}}
\def\eeq{\end{equation}}
\def\vk{{\bf k}}

\begin{document}

%Fill out the title section
\title{Two-site fluctuations and multipolar inter-site exchange interactions in strongly correlated systems}

%Author information
\author{L.~V.~Pourovskii}
\affiliation{Centre de Physique Th\'eorique, \'Ecole Polytechnique, CNRS, Universit\'e Paris-Saclay, 91128 Palaiseau, France}
\affiliation{Coll\`ege de France, 11 place Marcelin Berthelot, 75005 Paris, France}
\affiliation{Materials Modeling and Development Laboratory, National University of Science and Technology "MISIS", Moscow, Russia}

\begin{abstract}
An approach is proposed for evaluating dipolar and multipolar inter-site interactions in strongly-correlated materials. This approach is based on the single-site dynamical mean-field theory (DMFT) in conjunction with the atomic   approximation for the local self-energy. Starting from the local-moment paramagnetic state described by DMFT we derive inter-site interactions by considering the response of the DMFT grand potential to small  fluctuations of atomic configurations on two neighboring sites. The present method is validated by applying it to one-band and two-band $e_g$ Hubbard models on the simple-cubic 3$d$ lattice. It is also applied to study the spin-orbital order in the parent cubic structure of ternary chromium fluoride KCrF$_3$. We obtain the onset of a G-type antiferro-orbital order at a significantly lower temperature compared to that in real distorted KCrF$_3$. In contrast, its layered A-type  antiferromagnetic order and N\'eel temperature are rather well reproduced. The calculated full Kugel-Khomskii Hamiltonian contains spin-orbital coupling terms inducing a misalignment in the antiferro-orbital order upon the onset of antiferromagnetism.

\end{abstract}

\maketitle

\section{Introduction}

Magnetic and orbital-ordering phenomena in strongly-correlated materials have been a hot topic in condensed matter research for many years. In particular, transition-metal (TM) oxides and fluorides have attracted a lot of attention due to a complex interplay of their spin and orbital orderings\cite{kugel_khomskii}. More recently a lot of research have been focused on the lanthanide and actinide compounds exhibiting ordering of high-ranking multipoles, e.g., CeB$_6$ \cite{Shiina1997,Kuramoto2009}, actinide dioxides AO$_2$ (A=U, Np, Pu)\cite{Santini2009}, and URu$_2$Si$_2$, where the nature of the "hidden-order" phase is still hotly debated \cite{Mydosh2011}. Experimental determination of multipole-ordered structures is a complicated task, because  the conventional neutron diffraction method is often not  applicable in this case \cite{Santini2009}.

Dipolar and multipolar moments in those materials are carried by localized shells of correlated $d$ and $f$ electrons. First-principles description of such strongly correlated compounds is nowadays possible using a combination\cite{anisimov97,Kotliar2006} of density-functional-theory (DFT) band structure techniques with the  dynamical mean-field theory (DMFT) treatment\cite{Georges1996} of correlated electrons. This approach is particularly efficient in capturing the high-temperature symmetry-unbroken state. There are no principal limitations for applications of the same method to symmetry-broken ordered phases. However, typical low ordered temperatures and low symmetries of those phases as well as a vast configuration space of possible ordered states render  direct predictive DFT+DMFT calculations in this case rather difficult. Moreover, the single-site DMFT method suffers from the usual mean-field drawbacks overestimating ordering temperatures, especially, for  low-dimensional systems (see, for example, Ref.~\onlinecite{Rohringer2011,Hirschmeier2015,Maier2005,Schafer2015,Ayral2015,Horvat2015}). 

Hence, a promising approach for a first-principles description of orbital and multipolar ordering phenomena consists in using the DFT+DMFT method to evaluate an effective low-energy Hamiltonian describing inter-site interactions between localized shells. Such Hamiltonians can then be solved by a variety of methods developed for Heisenberg and similar models in order to predict the ordered phase as a function of external parameters like pressure or temperature. 

Several such techniques have been proposed \cite{Prange1979,Wang1982,Oguchi1983,Oguchi1983prb,Liechtenstein1987,Bruno2003,Ruban2004} for evaluating  low-energy spin Hamiltonians in conjunction with standard DFT methods. They were subsequently also extended to calculations, for example, of the magnetic crystalline anisotropy \cite{Solovyev1995} and Dzyaloshinskii-Moriya interactions\cite{Solovyev1996}  . Generally, in those approaches effective exchange interactions are extracted by considering a first-order response of the grand potential upon a simultaneous change of magnetic configurations of two neighbouring sites. In particular in those approaches that  have been to date  generalized for correlated systems  (Refs.~\onlinecite{Katsnelson2000,Pi2014,Secchi2015}), one computes the variation of the grand potential of a magnetically-ordered state upon simultaneous small tilting of two neighbouring spins.

The technique proposed in this work in order to calculate inter-site dipolar and multipolar interactions is similar in spirit to those methods. However, in contrast to them, we will calculate the variation of the DFT+DMFT grand potential of a {\it paramagnetic} (symmetry-unbroken) phase upon simultaneous small change of the atomic configurations of correlated shells of two neighboring sites. Hence, one can derive inter-site interactions directly from the high-temperature paramagnetic state, which is typically  most readily accessible for DFT+DMFT.   The approach is currently formulated using the atomic (Hubbard-I)\cite{hubbard_1} approximation for the DMFT local self-energy.  It is fast and, in principle, able to calculate all terms of the low-energy Hamiltonian, including non-Ising spin-spin, spin-orbital and  multipolar interactions. The formulation on the basis of Hubbard-I entails, however, certain limitations. In particular, the present approach is suitable for localized systems like TM oxides and local-moment lanthanide compounds and cannot be applied to metals. 

As first application of this technique to real materials we study the spin-orbital ordering in the cubic phase of the Mott insulator KCrF$_3$. In this compound the 3$d$ shell of the Cr$^{2+}$ ion is in the high-spin $t_{2g}^3e_g^1$ configuration with the spin of single $e_g$ electron aligned to that of the half-filled $t_{2g}$ subshell by the Hund's rule coupling, similarly to undoped peroxide manganese LaMnO$_3$.   KCrF$_3$ adopts the cubic peroxide structure at high temperatures. At $T_{OO}\approx 973$~K it undergoes a first-order orbital-ordering transition accompanied by a tetragonal distortion (space group $I4/mcm$)\cite{Margadonna2007}. Another structural transition to a low-temperature monoclinic phase (space group $I112/m$) due to tilting of the CrF$_6$ octahedra is observed at $T\approx 250$~K\cite{Margadonna2006}. Finally, a transition into an incommensurate layered antiferromangetic (AFM) phase with the ordering vector (1/2$\pm \delta$, 1/2$\pm \delta$, 0) in the monoclinic cell is taken place at $T_{N}\approx 80$~K\cite{Xiao2010}. Below 46~K the AFM order becomes a fully commensurate A-type one with $\delta \rightarrow 0$,  a spin canting is detected below 9.5~K leading to formation of a small ferromagnetic moment \cite{Xiao2010}. 

As in other Jahn-Teller systems it is important to disentangle the lattice and purely electronic superexchange contributions into the spin-orbital ordering in KCrF$_3$ to understand their relative importance. Previously the orbital ordering in the undistorted cubic structure has been studied theoretically within DFT+DMFT\cite{Autieri2014} and DFT+U\cite{Giovannetti2008,Xing2014} approaches. In particular, the authors of Ref.~\onlinecite{Autieri2014} derived an effective DMFT impurity problem for the Cr $e_g$ subshell with a simplified treatment of its interactions with the $t_{2g}$ spin, which was subsequently solved by a quantum Monte Carlo (QMC) method. They obtained a substantially underestimated value $T_{OO}\approx 400$~K  when only the supexchange contribution was taken into account.

Here we compute all relevant superexchange interactions for the cubic phase of KCrF$_3$ and then solve the resulting effective spin-orbital Hamiltonian within mean-field obtaining orbital and magnetic ordering temperatures and the corresponding phases. We find an underestimated value of $T_{OO}$ in agreement with Ref.~\onlinecite{Autieri2014}, in contrast, the calculated value for $T_{N}$ and the predicted A type of the AFM order agree with those experimentally observed in KCrF$_3$. We show that the onset of the AFM phase produces a feedback effect on the orbital arrangement leading a loss of the perfect antiferro-orbital order even in the absence of lattice distortions. 

The rest of  paper is organized as follows: the method is derived in Sec.~\ref{method_sec}.  It is subsequently tested and its limitations explored by applying it to one-band  and  two-band $e_g$ Hubbard model on the simple-cubic lattice in Secs.~\ref{hub_1b} and \ref{hub_2b}, respectively. Finally, its application to KCrF$_3$ and the obtained results are presented in Sec.~\ref{KCrF3_sec}.

\section{Method}\label{method_sec}
%Method

We start by deriving in Sec.~\ref{local_fluct} variation of the Hubbard-I self-energy with respect a change of the atomic configuration of correlated shell . The derived expressions are then used in Sec.~\ref{sec_grand_pot} to calculate the variation of the DFT+DMFT grand potential upon simultaneous change of atomic configurations on two neighboring sites and, thus, to extract the corresponding inter-site interactions between those configurations. Finally, in Sec.~\ref{sec_multipolar} we recast the obtained interactions into a more conventional dipolar and multipolar form. The full calculational procedure is shortly outlined in Sec.~\ref{calc_outline}.

\subsection{Local fluctuations within the Hubbard-I approximation}\label{local_fluct}

Let us first outline the main features of the Hubbard-I approximation (HIA) as applied  to the DMFT quantum impurity problem. In this case the HIA can be derived by a high-frequency expansion of the DMFT self-consistency condition (see, e.g., Ref.~\onlinecite{Pourovskii2007}) to the first order in $1/\omega$, leading to the following expression for the non-interacting level positions of the impurity:

\beq\label{eq_levpos}
\epsilon=-I\mu+\sum_{\vk} P_{\vk} H_{KS}^{\vk} P^{\dagger}_{\vk} - \Sigma_{dc},
\eeq
where $H_{KS}^{\vk}$ and $P_{\vk}$ are the Kohn-Sham(KS) Hamiltonian and "projector" between the KS and correlated spaces for a given $\vk$ point in the Brillouin zone (BZ), respectively, $\Sigma_{dc}$ is the double-counting correction for the self-energy, $\mu$ is the chemical potential, $I$ is the unit matrix. The  DMFT bath Green's function ${\cal G}$ within the HIA takes a very simple form

\beq\label{eq_Gbath}
{\cal G}^{-1}_0(i\omega_n)=i\omega_n I-\epsilon,
\eeq
where $\omega_n=\pi T (2n-1)$ is the fermionic Matsubara frequency. Solving of the impurity problem is then reduced to the diagonalization of the effective atomic Hamiltonian $H_{at}=\sum_{ab}\epsilon_{ab}f^{\dagger}_a f_b+H_U$, where $f^{\dagger}_a$($f_b$) is the creation(annihilation)  operator for the localized orbital labeled by relevant quantum numbers designated by $a$($b$), $H_U$ is the on-site Coulomb repulsion.

The corresponding atomic Green's function then reads 
\beq\label{eq_Gat}
G^{at}_{ab}(i\omega_n)=\sum_{\gamma\gamma'}\frac{\langle \gamma|f_a|\gamma'\rangle \langle \gamma'|f^{\dagger}_b|\gamma\rangle}{i\omega_n-E_{\gamma'}+E_{\gamma}}(X_{\gamma}+X_{\gamma'}),
\eeq
where $|\gamma\rangle$ and $|\gamma'\rangle$  are eigenstates of the atomic Hamiltonian $\hat{H}_{at}$, $E_{\gamma}$ and $X_{\gamma}=\frac{e^{-\beta E_{\gamma}}}{Z}$ are the corresponding eigenenergies and Boltzmann weights, respectively, $Z$ is the partition function, $\beta=\frac{1}{T}$ is the inverse temperature. The atomic self-energy can then be calculated through the Dyson equation:
\beq\label{Dyson}
\Sigma^{at}(i\omega_n)=[{\cal G}_0(i\omega_n)]^{-1}-[G^{at}(i\omega_n)]^{-1}
\eeq

In cases where the HIA is applicable and for reasonable temperatures 
%one may safely neglect $X_{\Gamma}$ for states not belonging to the ground state multiplet 
the system is far from the intermediate-valence regime, hence, charge fluctuations can be safely neglected. Moreover, in localized systems the most important fluctuations are expected to occur among quasi-degenerate states belonging to the ground-state (GS) atomic multiplet.
For example, for 4$f$ shells this multiplet is defined by the occupancy as well as by the spin $S$, orbital $L$ and total $J$ quantum numbers, in TM ions it is rather defined by the occupancy, $S$, and crystal field. In solids the GS multiplet can be additionally split by smaller energy scales, like the crystal field in rare-earths and the spin-orbit coupling in TM ions. Hence, here we consider fluctuations only among the states belonging to GS multiplet. It is useful for the following to recast the atomic GF into a slightly more general form:
\beq\label{eq_Gat_mat}
G^{at}=Tr\left[\hat{\rho} \hat{G}\right]+G^{at}_{1},
\eeq
where the first term comprises all contributions to $G^{at}$ involving the states of the GS multiplet; those states will be in the following designated by capital Greek letters, for example, $|\Gamma\rangle$. The rest is collected in $G^{at}_{1}$.   The density matrix $\hat{\rho}$ (throughout Sec.~\ref{method_sec} we use the hat, $\hat{X}$, for any matrix $X$ in the basis of atomic states $|\Gamma\rangle$) of the GS multiplet in the symmetry-unbroken (paramagnetic) state is defined within the HIA by

\beq\label{rho_at}
\rho_{\Gamma\Gamma'}=\delta_{\Gamma\Gamma'}\frac{e^{-\beta E_{\Gamma}}}{Z},
\eeq
where $\delta_{\Gamma\Gamma'}$ is the Kronecker delta,
and the corresponding element of the atomic GF matrix $\hat{G}$ in the imaginary time domain reads $G^{\Gamma\Gamma'}_{ab}(\tau)=-\left\langle\Gamma| T[f_a(\tau) f_b^{\dagger}(0)]|\Gamma'\right\rangle$, where $T$ is the time-ordering operator, $\left|\Gamma\right\rangle$ and $\left|\Gamma'\right\rangle$ are eigenstates of $H_{at}$ belonging to the GS multiplet. By the Fourier transform one obtains, e.g., for the off-diagonal matrix elements of $ \hat{G}$ in the frequency space:

\begin{widetext}
\beq\label{G_off_iw}
G^{\Gamma\Gamma'}_{ab}(i\omega_n)=\sum_{\lambda \in Q+1} \frac{1+e^{-\Delta E_{\Gamma\lambda}\beta}}{i\omega_n-\Delta E_{\Gamma\lambda}}(F^a)_{\Gamma\lambda}(F^{b\dagger})_{\lambda\Gamma'}+\sum_{\lambda \in Q-1} \frac{1+e^{-\Delta E_{\Gamma'\lambda}\beta}}{i\omega_n+\Delta E_{\Gamma'\lambda}}(F^{b\dagger})_{\Gamma\lambda}(F^a)_{\lambda\Gamma'},
\eeq
\end{widetext}
where $(F^{a(\dagger)})_{\Gamma\lambda}=\left\langle\Gamma|f_a^{(\dagger)}|\lambda\right\rangle$,
$\Delta E_{\Gamma\lambda}=E_{\lambda}-E_{\Gamma}$ is the energy difference between the state $\left|\Gamma\right\rangle$ belonging to the GS multiplet with the  occupancy $Q$ and the excited state $\left|\lambda\right\rangle$. Similar, but simpler expressions can be obtained for the diagonal elements $G^{\Gamma\Gamma}_{ab}$.

Let us now consider the change of the atomic Green's function upon a small fluctuation of the density matrix $\hat{\rho}$ with respect to its symmetry-unbroken Hubbard-I form (\ref{rho_at}). We define the fluctuation for diagonal elements $\rho_{\Gamma\Gamma}$ as a diagonal  $N \times N$ matrix $\delta\hat{\rho}^{\Gamma\Gamma}$ with the following elements:

\beq\label{diag_fluct}
\delta\rho^{\Gamma\Gamma}_{\Lambda\Lambda}=\left(\frac{N-1}{N}\delta_{\Lambda\Gamma}+\frac{1}{N}(\delta_{\Lambda\Gamma}-1)\right)\epsilon,
\eeq
where  $N$ is the degeneracy of the ground-state multiplet, $\epsilon$ is a small parameter. As one may easily see, the fluctuation (\ref{diag_fluct}) conserves the trace of $\hat{\rho}$ and induces a corresponding fluctuation of an angular moment of the shell. For example, if in the symmetry-unbroken state the value of an angular moment operator $\hat{J}$ is zero, $Tr\left[\hat{\rho} \hat{J}\right]=0$ and $\left\langle\Gamma|\hat{J}|\Gamma\right\rangle=J_{\Gamma}$, then the corresponding  fluctuation of the moment is $Tr\left[\delta\hat{\rho}^{\Gamma\Gamma} \hat{J}\right]=\epsilon J_{\Gamma}$.

We also define the off-diagonal fluctuation of $\delta\hat{\rho}^{\Gamma\Gamma'}$ as an $N \times N$ matrix with a single none-zero element:
\beq\label{off_diag_fluct}
\delta\rho^{\Gamma\Gamma'}_{\Lambda\Lambda'}=\delta_{\Gamma\Lambda}\delta_{\Gamma'\Lambda'}\epsilon .
\eeq 

Using the definition (\ref{eq_Gat_mat}) for the atomic GF one then obtains the following expression for the variational derivative of $G_{at}$ over a fluctuation of the type (\ref{diag_fluct}) or (\ref{off_diag_fluct}):

\beq\label{Gat_fluct}
\frac{\delta G^{at}}{\delta\hat{\rho}^{\Gamma\Gamma'}}=G^{\Gamma'\Gamma}-\delta_{\Gamma\Gamma'}\frac{Tr[\hat{G}]}{N}.
\eeq

The second term in (\ref{eq_Gat_mat}), $G^{at}_{1}$, does not contribute to the variational derivative (\ref{Gat_fluct}), because the weights  $X_{\gamma}$ of the states not belonging to the GS multiplet are not affected by fluctuations of the types (\ref{diag_fluct}) and (\ref{off_diag_fluct}). Those fluctuations only redistribute the weights within the GS multiplet and do not change $Z$. 

The corresponding variational derivative of the atomic self-energy (\ref{Dyson}) reads:
\beq\label{Self_fluct}
\frac{\delta \Sigma^{at}}{\delta\hat{\rho}^{\Gamma\Gamma'}}=[G^{at}]^{-1}\left(G^{\Gamma'\Gamma} - \delta_{\Gamma\Gamma'}\frac{Tr[\hat{G}]}{N}\right) [G^{at}]^{-1}
\eeq

In the next section we will make use of (\ref{Self_fluct}) to calculate a response of the DFT+DMFT grand potential upon small fluctuations of the density matrix (\ref{rho_at}) on two neighbouring sites.

\subsection{Response of the grand potential and effective inter-site interactions}\label{sec_grand_pot}

The DFT+DMFT grand potential\cite{Savrasov2004,Georges2004,Kotliar2006} reads 

\begin{widetext}
\begin{eqnarray}\label{dmft_gp}
\Omega\left[n({\bf r}),G^{loc},\Delta \Sigma,V_{KS}\right]=-\frac{1}{\beta}Tr \ln\left[i\omega_n+\mu+\frac{\nabla^2}{2}-V_{KS}- \Delta\Sigma
\right]  -Tr\left[G^{loc}\Delta \Sigma\right]+ \\
  \sum_{{\bf R}}\left[\Phi^{imp}[G^{loc}_{\bf R}]-\Phi^{dc}[G^{loc}_{\bf R}]\right]+\Omega_{r}[n({\bf r})] \equiv\Delta \Omega\left[G^{loc},\Delta \Sigma,V_{KS}\right]+\Omega_{r}[n({\bf r})] \nonumber ,
\end{eqnarray}
\end{widetext}
where $n({\bf r})$ is the electronic density, $V_{KS}$ is the Kohn-Sham one-electron potential, $G^{loc}$ is the local GF, 
$\Delta\Sigma$ is the difference between the impurity self-energy $\Sigma^{imp}$ and the double counting correction
 $\Sigma^{dc}$, $\Phi^{imp}[G^{loc}_{\bf R}]$ is the 
DMFT interaction energy functional for the site 
${\bf R}$,  
 $\Phi^{dc}[G^{loc}_{\bf R}]$ is the corresponding functional for the double-counting correction, $\mu$ is the chemical potential. The last term $\Omega_{r}[n({\bf r})]$ depends only on the electronic charge density $n({\bf r})$, while all other terms collected in $\Delta \Omega\left[G^{loc},\Delta \Sigma,V_{KS}\right]$ do not have an explicit dependence on $n({\bf r})$. At the DMFT self-consistency  the local GF of the lattice problem $G^{loc}$ should be equal to the impurity GF $G^{imp}$. 
%Hence, the expression for DMFT grand potential (\ref{dmft_gp}) is often writen as a functional of $G^{loc}$. 
Within the HIA, however, the full DMFT  self-consistency is never achieved because the hybridization function is neglected in the impurity problem, $G^{imp}\equiv G^{at}$, but is included into the local GF of the lattice problem. Hence, within the HIA one should always keep the distinction between $G^{loc}$ and $G^{imp}$, where $G^{imp}$ and $\Sigma^{imp}$ calculated within the HIA in accordance with (\ref{eq_Gat}) and (\ref{Dyson}), respectively.

Let us now introduce the basis of Kohn-Sham eigenstates $\{\Psi_{\vk\nu}\}$, where $\nu$ labels Kohn-Sham bands. The corresponding real-space (Wannier) basis functions are  defined by  $\Psi_{{\bf R}\nu}({\bf r}-{\bf R})=\frac{V}{(2\pi)^3}\int_{BZ} d \vk e^{-i{\bf kR}}\Psi_{\vk\nu}({\bf r})$, where $V$ is the unit cell volume \footnote{The Wannier transformation is gauge-invariant with respect to a unitary transformation of $\{\Psi_{\vk\nu}\}$. For the present derivation it is not important, hence, we assume that the corresponding unitary matrix is equal to unity.}. We also introduce a real-space basis of (localized) Wannier orbitals representing correlated states, $\{w_{{\bf R}a}\}$, where $a$ labels orbitals at the correlated shell ${\bf R}$, as well as corresponding projectors between the KS and correlated subspaces, $P^{\bf RR'}_{a\nu}=\left\langle w_{{\bf R}a}|  \Psi_{{\bf R'}\nu}\right\rangle$. Using the real-space bases  $\{\Psi_{{\bf R}\nu}\}$ and $\{w_{{\bf R}a}\}$ and within the HIA one may rewrite $\Delta \Omega$ as follows:

\begin{widetext}
\beq\label{d_Om}
\Delta \Omega\left[G^{loc},\Delta\Sigma,V_{KS}\right]=-\frac{1}{\beta} Tr \ln [\mathcal{M}_n]-\sum_{\bf R} Tr\left[G^{loc}_{\bf R}\Sigma^{at}_{\bf R}\right] + \sum_{\bf R} Tr\left[G^{loc}_{\bf R}\Sigma^{dc}_{\bf R}\right] + \sum_{\bf R} \left[\Phi^{at}[G^{loc}_{\bf R}]-\Phi^{dc}[G^{loc}_{\bf R}]\right],
\eeq
\end{widetext}
where elements of the real-space matrix $\mathcal{M}_n$ read

%\begin{eqnarray}
\begin{align}\label{M}
\mathcal{M}^{\bf RR'}_n= & (i\omega_n+\mu)I-H_{KS}^{\bf RR'} \\
& - \sum_{\bf R''}P^{\dagger}_{\bf RR''}(\Sigma^{at}_{ \bf R''R''}(i\omega_n)-\Sigma^{dc}_{ \bf R''R''})P_{\bf R'' R'}, \nonumber
%\end{eqnarray}
\end{align}
$H_{KS}^{\bf RR'}$  and $\Sigma^{at}_{ \bf RR}(i\omega_n)$ are matrices in the band and correlated orbitals' spaces, respectively, the matrix elements of the former are given by $\left[H_{KS}^{\bf RR'}\right]_{\nu\nu'}=\langle \Psi_{{\bf R}\nu}|-\frac{\nabla^2}{2}+V_{KS}|\Psi_{{\bf R'}\nu'}\rangle $.

We will now calculated the response of the grand potential (\ref{dmft_gp}) upon simultaneous fluctuations of atomic configurations of correlated shells on two {\it different} atomic sites, i.e. we evaluate $\frac{\delta^2\Omega}{\delta \hat{\rho}^{\Gamma_1\Gamma_2}({\bf R})\delta  \hat{\rho}^{\Gamma_3\Gamma_4}({\bf R'})} $. First, in the usual "force theorem" spirit \cite{force_theorem,Liechtenstein1987,Solovyev1998}   one may neglect, to the first order in $\delta \hat{\rho}^{\Gamma_1\Gamma_2}({\bf R})\delta  \hat{\rho}^{\Gamma_3\Gamma_4}({\bf R'})$, the contribution due to the renormalization of the charge density, i.e., the contribution from $\Omega_{r}[n({\bf r})]$. One may also notice that all terms in (\ref{d_Om}), apart from the first one, are site-diagonal and will not contribute to a variational derivative over configurations of two different sites. Hence, the only non-zero contribution due to simultaneous fluctuations on two different sites ${\bf R}$ and ${\bf R'}$ is due to the first term in (\ref{d_Om}). $\mathcal{M}_n$ dependence on the correlated shell configuration stems from that of the atomic self-energy $\Sigma^{at}$. The double-counting correction $\Sigma^{dc}$ for a paramagnetic phase depends only on the total shell occupancy, which is not affected by the density-matrix variations (\ref{diag_fluct}) and (\ref{off_diag_fluct}). Performing the derivative  $\frac{\delta^2\left[-\frac{1}{\beta} Tr \ln [\mathcal{M}_n] \right]}{\delta  \hat{\rho}^{\Gamma_1\Gamma_2}({\bf R})\delta  \hat{\rho}^{\Gamma_3\Gamma_4}({\bf R'})}$  and making use of the "folding" property of projector matrices, $\sum_{\bf R_1R_2}P_{\bf RR_1} \left[\mathcal{M}_n^{-1}\right]_{\bf R_1R_2}\left[P_{\bf R_2R'}\right]^{\dagger}=G_{\bf RR'}$  one obtains

\begin{widetext}
\beq\label{V}
\frac{\delta^2\Omega}{\delta  \hat{\rho}^{\Gamma_1\Gamma_2}({\bf R})\delta  \hat{\rho}^{\Gamma_3\Gamma_4}({\bf R'})}\equiv \langle M_1 M_3| V^{\bf RR'}| M_2 M_4\rangle=\frac{1}{\beta} Tr \left[ G_{\bf RR'}\frac{\delta\Sigma^{at}_{\bf R'}}{\delta \rho^{\Gamma_3\Gamma_4}} G_{\bf R'R}\frac{\delta\Sigma^{at}_{\bf R}}{\delta \rho^{\Gamma_1\Gamma_2}}\right],
\eeq
where the derivative $\frac{\delta\Sigma^{at}_{\bf R}}{\delta \hat{\rho}^{\Gamma_1\Gamma_2}}$ over an on-site fluctuation is given by eq. (\ref{Self_fluct}), $M_l$ etc.  is the relevant set of quantum numbers labeling the state $\Gamma_l$ and the "inter-site" GF $G_{\bf RR'}$ can be calculated as a Fourier transform of the DMFT lattice GF in the reciprocal space:

\beq\label{G_off}
G_{\bf RR'}(i\omega_n)=\frac{V}{(2\pi)^3}\int_{BZ} d \vk e^{-i{\bf k}({\bf R'}-{\bf R})} P_{\vk}\left[i\omega_n+\mu-H_{KS}^{\vk}-P_{\vk}^{\dagger}\Delta\Sigma P_{\vk}\right]^{-1} P_{\vk}^{\dagger}.
\eeq
\end{widetext}

In eq. (\ref{V}) we identify $\frac{\delta^2\Omega}{\delta \hat{\rho}^{\Gamma_1\Gamma_2}({\bf R})\delta \hat{\rho}^{\Gamma_3\Gamma_4}({\bf R'})}$ with the corresponding inter-site interaction of an effective low-energy Hamiltonian of the system:

\beq\label{H_eff}
\hat{H}_{eff}=\sum_{{\bf R},\Gamma}E_{\Gamma} \hat{\rho}^{\bf R}_{\Gamma\Gamma}+\sum_{\substack{{\bf RR'} \\ 1 2 3 4}} \langle 13| V^{\bf RR'}|  24\rangle \hat{\rho}^{\bf R}_{12} \hat{\rho}^{\bf R'}_{34},
\eeq
where $\hat{\rho}^{\bf R}_{\Gamma\Gamma_1}=|\Gamma^{\bf R}\rangle \langle \Gamma_1^{\bf R}|$ is the corresponding projection (Hubbard) operator between the atomic states $\Gamma$ and $\Gamma_1$ belonging to the ground-state multiplet of the site ${\bf R}$, $E_{\Gamma}$ is the one-site (crystal-field) term, $ \langle 13| V^{\bf RR'}|  24\rangle$ is the inter-site interaction between the corresponding Hubbard operators on the sites ${\bf R}$ and ${\bf R'}$ (here the label $\Gamma$ is suppressed and the short-hand notation $1\equiv M_1$ is used). 

The identification of the corresponding inter-site interaction in $\hat{H}_{eff}$ with (\ref{V}) can be justified using, e.g., the approach of Refs.~\onlinecite{Plefka1982,Georges1991,Georges2004}. Using this approach one may write a (Legendre-transformed) grand potential corresponding to (\ref{H_eff}) for a set of preassigned on-site occupancy matrices $\{ \hat{\rho}^{\bf R}\}$ as $\Omega_{LT}[\rho]=\Omega_0[\rho]+\sum_{\substack{{\bf RR'}\\ 1234}}\langle 13| V^{\bf RR'}|  24\rangle  \hat{\rho}^{\bf R}_{12}  \hat{\rho}^{\bf R'}_{34}+\Omega_{corr}=\Omega_{MF}+\Omega_{corr}$, where $\Omega_{MF}$ and $\Omega_{corr}$ is the mean-field  and beyond-mean-field contributions, respectively, $\Omega_0[\rho]$ is the one-site term. Setting the density matrices $\hat{\rho}^{\bf R}$ to their mean-field values in the symmetry-unbroken state and computing the variational derivative of $\Omega_{MF}$ over $\delta  \hat{\rho}^{\bf R}_{12} \delta  \hat{\rho}^{\bf R'}_{34}$  one obtains $\langle 13| V^{\bf RR'}|  24\rangle$. Hence, one identifies the derivative (\ref{V}) of the dynamical {\it mean-field} grand potential (\ref{dmft_gp}) as the corresponding inter-site interaction in (\ref{H_eff}). Of course, the usefulness of those interactions depends on whether the effective Hamiltonian (\ref{H_eff}) indeed describes the low-energy physics of (\ref{dmft_gp}). This should be the case for strongly-correlated local-moment systems, e. g., for  rare-earth intermetallics above their Kondo temperature or for Mott insulators.

\subsection{Multipolar formalism}\label{sec_multipolar}

The inter-site interactions between atomic states $|\Gamma\rangle$ calculated in accordance with (\ref{V}) can be used directly, e.g., in an effective Hamiltonian  of the type (\ref{H_eff}). This Hamiltonian  is written in terms of low-energy interactions between the on-site Hubbard operators defined above describing transitions between atomic states belonging to the ground-state multiplet.

However, the standard dipolar and multipolar tensor operators are, in fact, linear combinations of those Hubbard operators with coefficients written in terms of of the  corresponding Wigner 3$j$ symbols\cite{Santini2009,Blum_DM}. Hence, instead of working directly with the Hubbard-operator form  (\ref{H_eff})   one may recast this Hamiltonian to describe interactions between dipole and multipole (quadrupole, octopole, etc.) operators acting on neighboring sites. The low-energy Hamiltonian  in this form is more standard (one may recall, for example, the spin Heisenberg and spin-orbit Kugel-Khomskii Hamiltonians) and also more compact when additional symmetries are present. Moreover, it is written in terms of operators which expectation values , i. e. dipole and multipole moments, are directly  measured experimentally. In this section we  derive a transformation relating inter-site interactions in the density-matrix (\ref{H_eff}) and more conventional dipolar-multipolar Hamiltonians.  

We start by briefly summarizing properties of tensor operators. The spherical tensor operators in the basis of angular-momentum eigenstates $|JM\rangle$  are standardly defined as follows \cite{Santini2009,Blum_DM}:
 
\begin{align}\label{T}
\hat{T}_{KQ}(J) &=\sum_{MM'} T^{MM'}_{KQ}(J)|JM\rangle \langle JM'| \\
 &= \sum_{MM'} T^{MM'}_{KQ}(J) \hat{\rho}_{MM'}, \nonumber
\end{align}
where $K$ and $Q$ label the multipole rank and component, respectively, $2J+1$ states $|JM\rangle$ belong to the ground-state multiplet specified by the angular-momentum quantum number $J$, $M=-J,..,J$, $\hat{\rho}_{MM'} \equiv |JM\rangle \langle JM'|$ is the Hubbard operator acting within the ground-state multiplet, the coefficients $T^{MM'}_{KQ}(J)$ read:

\beq\label{TMM}
T^{MM'}_{KQ}(J)=(-1)^{J-M}(2K+1)^{1/2}\left( \begin{array}{ccc}
J & J & K \\
M' & -M & Q \end{array} \right).
\eeq
The set of $(2J+1)^2$ 
operators $\hat{T}_{KQ}(J)$ with $K=0,1,...,2J$ , 
(i.e., monopole, dipole etc. operators) and $Q=-K,..,K$ is complete in the subspace spanned by the $|JM\rangle$ states and any operator acting in this subspace can be represented as a linear superposition of $\hat{T}_{KQ}(J)$. Other properties of those operators are discussed, e.g., in Refs.~\onlinecite{Santini2009,Blum_DM}. In particular, one may notice that the tensor operators (\ref{T}) cannot represent observables as they are not self-adjoint\cite{Blum_DM}, $\hat{T}_{KQ}^{\dagger}=(-1)^{-1}\hat{T}_{K-Q}$,  for $Q\neq0$ . However, the self-adjoint linear combinations of $\hat{T}_{KQ}$ can be formed similarly to the real spherical harmonics:

\beq\label{Oop}
\hat{O}_{KQ}(J)= \sum_{MM'} O^{MM'}_{KQ}(J) \hat{\rho}_{MM'},
\eeq
where 
\begin{eqnarray}\label{Omm}
O^{MM'}_{KQ}(J) =\frac{1}{\sqrt{2}}[(-1)^Q T^{MM'}_{KQ}(J)+T^{MM'}_{K-Q}(J)] \\  
O^{MM'}_{KQ}(J) =\frac{i}{\sqrt{2}}[ T^{MM'}_{K-Q}(J)-(-1)^QT^{MM'}_{KQ}(J)]\nonumber,
\end{eqnarray}
 for $Q >0$ and $Q<0$, respectively. For example, for the dipole, $K=1$, the components $Q$ equal to $-1$, $0$, and $1$ transform under rotations as Cartesian  $y$, $z$, and $x$, respectively, similarly to the corresponding real spherical harmonics.

One may introduce inter-site interaction between the tensor operators (\ref{Oop}) acting at sites ${\bf R}$ and  ${\bf R'}$   as $\sum_{\substack{KK' \\ QQ'}} V^{QQ'}_{KK'}({\bf RR'}) \hat{O}_{KQ}({\bf R})\hat{O}_{K'Q'}({\bf R'})$, where the tensor operators of  the rank $K$($K'$) and for the component $Q$($Q'$) are defined for the ground-state multiplet $J$ of the corresponding atomic shell ${\bf R}$(${\bf R'}$), respectively.  The label $J$ in $\hat{O}_{KQ}({\bf R})$ is suppressed here to simplify the notation. It is easy to show that $\sum_{\substack{KK' \\ QQ'}} V^{QQ'}_{KK'}({\bf RR'}) O_{KQ}^{M_1M_2} O_{K'Q'}^{M_3M_4}$ is equal to   the inter-site interaction $\langle M_1 M_3| V^{\bf RR'}| M_2 M_4\rangle$ defined in (\ref{V}) and (\ref{H_eff}).

By making use of the orthogonality relations of the 3$j$-symbols one may also show that $\sum_{MM'}O^{MM'}_{KQ}(J)O^{M'M}_{K'Q'}(J)=\delta_{KK'}\delta_{QQ'}$. Then by multiplying the inter-site interactions (\ref{V}) by $O_{KQ}^{M_2M_1}(J)$ and $O_{KQ}^{M_4M_3}(J)$ and summing over the quantum numbers $M$ one obtains  
\begin{widetext}
\beq\label{V_to_mult}
\sum_{\substack{M_1M_2 \\ M_3M_4}} \langle M_1M_3| V^{\bf RR'}|  M_2M_4\rangle O_{KQ}^{M_2M_1}(J) O_{K'Q'}^{M_4M_3}(J)  = V^{QQ'}_{KK'}({\bf RR'}).
\eeq
\end{widetext}

Using (\ref{V_to_mult}) one may transform the inter-site interactions from the atomic-level, eq. (\ref{V}), to mutipolar form \footnote{One may notice that off-diagonal interactions  $\langle M_1M_3| V^{\bf RR'}|  M_2M_4\rangle$ with $M_1 \neq M_2$ and/or $M_3 \neq M_4$ may carry an arbitrary complex phase, which will then  be passed to $V^{QQ'}_{KK'}({\bf RR'})$. To avoid this we require that $|\Gamma\rangle \equiv |JM\rangle$ states used in (\ref{G_off_iw}) and (\ref{T}) satisfy the usual phase convention with the matrix element of the ladder operator $\langle JM|J_{+}|JM-1\rangle$ being a real positive number.}.

In many cases inter-site interactions and corresponding ordering temperatures come out to be much smaller than the crystal field (CF) splitting within the ground-state multiplet $J$. In this case one may restrict oneself to determining inter-site interactions between the states belonging to the lowest CF level. For example, one may represent the  state of an $e_g$ TM ion by a product of the ordinary spin $s$ and pseudo-spin $\tau$ quantum numbers, with the opposite directions of the pseudo-spin corresponding to the $3z^2-r^2$ and $x^2-y^2$ orbitals, respectively. This representation is widely used for TM oxides \cite{kugel_khomskii}. In this case one may introduce another type of tensor operators, the double tensor, which is a direct product of two spherical tensors for $J=1/2$ and can be written using (\ref{Omm}) as follows:

\begin{align}\label{O_ts}
\hat{O}^{\mu\nu}_{\Lambda\Sigma} & =\sum_{\tau\tau'}\sum_{ss'} O_{\Lambda\mu}^{\tau\tau'}(1/2) O_{\Sigma\nu}^{ss'}(1/2) |\tau s\rangle\langle \tau' s'| \\
 & =\sum_{\tau\tau'}\sum_{ss'}O^{\mu\nu}_{\Lambda\Sigma}(\tau s;\tau' s')\hat{\rho}_{\tau s;\tau' s'}, \nonumber
\end{align}
where we again use the corresponding Hubbard operator $\hat{\rho}_{\tau s;\tau' s'} \equiv |\tau s \rangle\langle \tau' s'|$.  The subscripts $\Lambda\Sigma$ and superscripts $\mu\nu$ in  $\hat{O}^{\mu\nu}_{\Lambda\Sigma}$  are the ranks and components, respectively, of the single tensors forming the direct product. Then, for example, in the "double" spin-orbital space $\hat{O}^{0p}_{01}=\hat{s}_p$ and $\hat{O}^{p0}_{10}=\hat{\tau}_p$ will designate the spin and orbital dipole tensors, respectively, with $p=x$, $y$, or $z$. Analogously, spin-orbital combined tensors can be  also introduced for the case of KCrF$_3$ or other compounds with the high-spin $t_{2g}^3e_g^1$ shell, the only difference is that the $J=2$ tensor $\hat{O}_{\Sigma\nu}(2)$ is used in this case to describe the spin.

 Similarly to (\ref{V_to_mult}) the corresponding interactions read

\beq\label{V_to_ts}
\sum_{1234} \langle 13| V^{\bf RR'}|  24\rangle O^{\mu\nu}_{\Lambda\Sigma}(2;1)O^{\mu'\nu'}_{\Lambda'\Sigma'}(4;3) = V^{\mu\nu;\mu'\nu'}_{\Lambda\Sigma;\Lambda'\Sigma'}({\bf RR'}),
\eeq
where the short-hand notation $1\equiv \{\tau_1 s_1\}$ is used. Finally, instead of using the tensors as defined in eqs.  (\ref{Oop}) and  (\ref{O_ts}) one may wish to write the interactions in terms of more conventional spin operators for the dipole case and the unit matrix  for the monopole one, respectively, by renormalizing the inter-site interactions as follows:

\beq\label{to_op}
J^{\mu\nu;\mu'\nu'}_{\Lambda\Sigma;\Lambda'\Sigma'}({\bf RR'})=V^{\mu\nu;\mu'\nu'}_{\Lambda\Sigma;\Lambda'\Sigma'}({\bf RR'})c(\Lambda)c(\Sigma)c(\Lambda')c(\Sigma'),
\eeq
where the factor $c(\Lambda)$ is equal to $\frac{1}{\sqrt{2}}$ and  $\sqrt{2}$ for $\Lambda$ equal to 0 and 1, respectively \cite{Blum_DM}. 

We will label those interaction by $\tau$ and  $s$ for the dipole orbital and spin moments, respectively, as well as by $q$ for the dipole-dipole ("quadrupole") spin-orbital one. For example,   $\hat{J}_{01;01}({\bf RR'})\equiv \hat{J}_{ss}({\bf RR'})$ is the spin-spin dipole-dipole interaction, $\hat{J}_{10;10}({\bf RR'}) \equiv \hat{J}_{\tau\tau}({\bf RR'}) $ is the orbital-orbital dipole-dipole one,  $\hat{J}_{10;11}({\bf RR'})\equiv \hat{J}_{\tau q}({\bf RR'})$ is the orbital-(spin-orbital) dipole-quadrupole one and so on. Finally, for the case of one atom per unit cell we may always make use of the translational invariance, hence, ${\bf RR'}$ can be substituted with $\Delta {\bf R} ={\bf R'}-{\bf R}$.

\subsection{Outline of calculation procedure}\label{calc_outline}

Let us summarize the sequence of steps for calculating inter-site interactions using the method described above. 

First, one carries out full self-consistent DFT+DMFT calculations using the Hubbard-I approximation as the impurity solver. Second, one computes the atomic Green's function matrix elements (\ref{G_off_iw}) and the variation derivative of the atomic self-energy (\ref{Self_fluct}) as well as the inter-site DMFT Green's function in the real-space (\ref{G_off}). Finally, the inter-site interactions are computed in accordance with (\ref{V}) and then transfomed, if desired, into a suitable multipolar form using (\ref{V_to_mult}) or (\ref{V_to_ts}). 
%The whole procedure is rather fast and easily parallelized, in results, inter-site interactions for tens of coordination shells can be calculated on a modest size computational cluster in several hours. 
The method is implemented  numerically  using the TRIQS library\cite{Parcollet2015}.  

\section{One-band Hubbard model}\label{hub_1b}

In this section we benchmark the approach presented in Sec.~\ref{method_sec} by applying it to a simple example of the one-band particle-hole symmetric Hubbard model on the 3$d$ simple cubic lattice. The Hamiltonian of this model reads

\begin{figure}
	\centering
	\includegraphics[width=0.85\columnwidth]{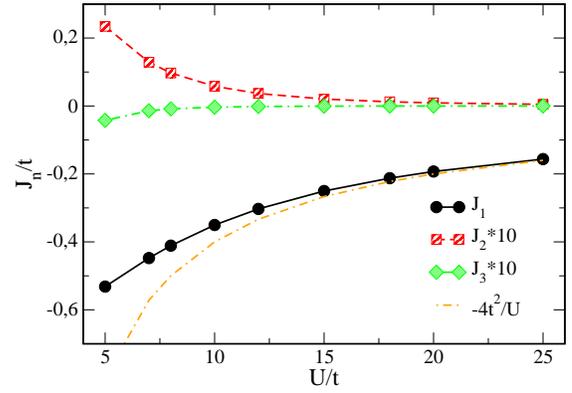}
	\caption{\label{fig:JvsU_1b} Calculated inter-site interactions $J_n$ for the first three coordinational shells. The values of  $J_2$ and $J_3$ are multiplied by 10. The dash-dotted line is the $-4t^2/U$ asymptote.}  
\end{figure}

\beq\label{1b_hub_h}
%\hat{H}_{1b}=\sum_{{\bf k}\sigma} \epsilon_{\bf k}f_{{\bf k}\sigma}^{\dagger}f_{{\bf k}\sigma}-\frac{U}{2}\sum_{i\sigma} \hat{n}_{i\sigma}+U\sum_{i}\hat{n}_{i\uparrow}\hat{n}_{i\downarrow},
\hat{H}_{1b}=\sum_{{\bf k}\sigma} \epsilon_{\bf k}f_{{\bf k}\sigma}^{\dagger}f_{{\bf k}\sigma}+U\sum_{i}\left(\hat{n}_{i\uparrow}-\frac{1}{2}\right)\left(\hat{n}_{i\downarrow}-\frac{1}{2}\right),
\eeq
where ${\bf k}$ belongs to the first Brillouin zone of the simple cubic lattice,  $\hat{n}_{i\sigma}=f_{i\sigma}^{\dagger}f_{i\sigma}$ is the number operator for the site $i$ and spin $\sigma$. For the simple cubic lattice with the nearest-neigbour hopping $t$ the band energy $\epsilon_{\bf k}=-2t(\cos k_x + \cos k_y + \cos k_z)$, where $k_{\alpha}$ are in units of the inverse lattice spacing $1/a$. Applying  the Hubbard-I approximation in the framework of DMFT to $\hat{H}_{1b}$ as described in Sec.~\ref{local_fluct} and  under the condition of $T \ll U$ one obtains 
$G^{at}(i\omega_n)=\left[\frac{1/2}{i\omega_n+U/2}+\frac{1/2}{i\omega_n-U/2}\right]$ and $\Sigma^{at}(i\omega_n)=\frac{U^2}{4 i\omega_n}$ for the atomic GF (\ref{eq_Gat}) and self-energy (\ref{Dyson}), respectively. Then one may easily obtain inter-site interactions of the effective low-energy model at $t \ll U$ analytically   by computing the inter-site GF using  the Fourier transform (\ref{G_off}) and the variational derivatives of the atomic self-energy using eqs. (\ref{G_off_iw}) and (\ref{Self_fluct}), respectively, and then inserting the result in (\ref{V}). For example, by inserting the nearest-neigbour inter-site GF $G_{{\bf RR'} \in NN}^{\sigma\sigma'}(i\omega_n)=-\frac{\delta_{\sigma\sigma'}t}{\left(i\omega_n-\frac{U^2}{4i\omega_n}\right)^2}$ and the "off-diagonal" derivative of the atomic self-energy ($\frac{\delta \Sigma^{at}}{\delta\hat{\rho}^{\downarrow\uparrow}}$ is the same expression transposed)
\begin{align}
	\frac{\delta \Sigma^{at}}{\delta\hat{\rho}^{\uparrow\downarrow}}(i\omega_n)= &   \left(i\omega_n-\frac{U^2}{4i\omega_n}\right)^2  
	\left( \begin{array}{cc} 
		0 & \frac{U}{\omega_n^2+U^2/4}  \\
		0 & 0 \end{array} \right)  
\end{align}
into (\ref{V}) and carrying out the summation over Matsubara frequencies and spins one obtains $\frac{2t^2}{U}$for the nearest-neighbor spin-off-diagonal matrix element $\langle \uparrow \downarrow| V({\bf d})| \downarrow \uparrow\rangle$ (where the lattice vector ${\bf d}={\bf R'}-{\bf R}$ connects nearest neighbors). This is indeed the correct value for this matrix element of the low-energy model for $\hat{H}_{1b}$ at $t \ll U$, which is well known to be the spin-1/2 Heisenberg model 

\beq\label{Hheis}
\hat{H}_{H}=-\sum_{ij}J_n \hat{s}_i\hat{s}_j,
\eeq
where the interaction is isotropic and depends only on the distance $|{\bf R}_i-{\bf R}_j|$, i.e. on the coordination shell $n$. In the lowest order in $t/U$ only the nearest neighbor anti-ferromagnetic interaction $J_1=-\frac{4t^2}{U}$ survives in $\hat{H}_{H}$.

We have calculated numerically all matrix elements of (\ref{Hheis}) for several first coordination shells as a function of U/t using (\ref{V}) and then applied the transformation (\ref{V_to_mult}) to obtain the corresponding inter-site interactions between the dipole tensor operators for  spin 1/2. As expected, those interactions come out to be isotropic and direction-independent, $V^{xx}_{11}({\bf d})=V^{yy}_{11}({\bf d})=V^{zz}_{11}({\bf d})=V_n$. Finally, the tensor interactions $V_n$ are renormalized, $J_n=2V_n$ (cf. eq.~\ref{to_op}), for the standard angular-momentum-operator form (\ref{Hheis}) of the Heisenberg Hamiltonian $\hat{H}_{H}$. Resulting  $J_n$ are plotted in Fig.~\ref{fig:JvsU_1b} as a function of $U/t$. One sees that $J_1$ deviates stronger from the $-4t^2/U$ asymptote with increasing $t$ and, simultaneously, the second and third coordination sphere interactions increase though they still remain quite insignificant compared to $J_1$.

\begin{figure}
	\centering
	\includegraphics[width=0.85\columnwidth]{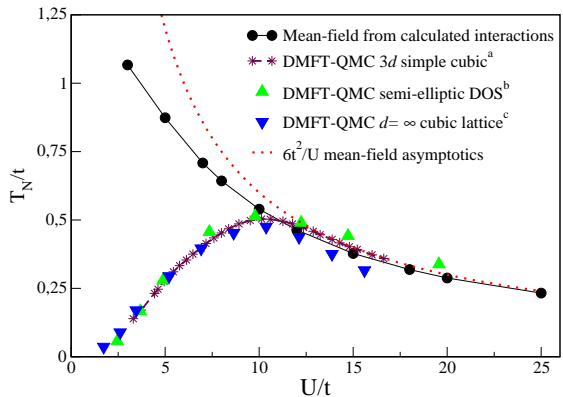}
	\caption{\label{fig:TNvsU_1b} Calculated values for the mean-field  N\'eel temperature $T_N$ compared with those obtained within QMC from a) Ref.~\onlinecite{Hirschmeier2015}  b) Ref.~\onlinecite{Ulmke1995}  c) Ref.~\onlinecite{Jarell1992} as well as with the large-$U$ asymptote $T_N=\frac{6t^2}{U}$. }  
\end{figure}

The calculated interactions $J_n$ have been used to evaluate the value of N\'eel temperature $T_N$ for the model (\ref{1b_hub_h})  within the mean-field approximation. The obtained values are compared in Fig.~\ref{fig:TNvsU_1b} to $T_N$ calculated within single-site DMFT using numerically-exact quantum Monte-Carlo (QMC) techniques \cite{Jarell1992,Ulmke1995,Hirschmeier2015}. The agreement with these numerically-exact $T_N$  is good for $U > 10t$. We note that within the dynamical mean-field theory Heisenberg $J_n$  not only define  the transition temperature $T_N$ but also directly impact spectral properties of the  N\'eel phase. In fact $J_n$ determine the spin-polaron peak structure within the Hubbard bands\cite{Rainer1992,Sangiovanni2006}, which can be in some cases detected in real Mott insulators\cite{Sangiovanni2006}. Hence, one may suggest that the present approach can be possibly used to provide parameters for $t-J$-like models aimed at investigating those phenomena. 

For  $U < 10t$ the present approach deviates significantly from the exact mean-field values, though less strongly than the simplest large-$U$ asymptote. The value of $U \approx 10t$ at which the maximum of exact mean-field $T_N$ is reached is very close to the critical value of $U$ for the metal-insulator transition in the paramagnetic phase \cite{Rozenberg1994}. Hence, one concludes that the present approach is reliable in the Mott-insulating regime.

\section{$e_g$-orbital Hubbard model}\label{hub_2b}

Here we apply the method of Sec.~\ref{method_sec} to a more complex model system, a two-band Hubbard model on the 3$d$ simple cubic lattice, given by

\beq\label{Hab}
\hat{H}_{2b}=\sum_{\substack{\langle ij\rangle \\ ab\sigma}} t_{ij}^{ab}f_{ia\sigma}^{\dagger}f_{jb\sigma} + \hat{H}_{int},
\eeq
where $\langle ij\rangle$ runs over nearest-neighbor bonds, $a$ and $b$ label orbitals, $t_{ij}^{ab}$ is the corresponding element of the hopping matrix, $\hat{H}_{int}$ is the on-site interaction term. We assume the orbitals to belong to the $e_g$ representation of the cubic group for $l=2$, $a\equiv 3z^2-r^2$ and $b\equiv x^2-y^2$, and employ the corresponding relations between the nearest-neighbor hopping integrals $t_{ij}^{ab}$, in which case the Fourier-transformed hopping matrix reads:

\begin{widetext}
\beq\label{2b_hoping}
t(\vk)=\left( \begin{array}{cc}
-\frac{1}{2}(\cos k_x + \cos k_y)-2 \cos k_z & \frac{\sqrt{3}}{2}(\cos k_x - \cos k_y) \\
\frac{\sqrt{3}}{2}(\cos k_x - \cos k_y) & -\frac{3}{2}(\cos k_x + \cos k_y) \end{array} \right)t,
\eeq
where $t$ is the hopping between two $3z^2-r^2$ orbitals for $\langle ij\rangle$ along the $\hat{z}$ axis. By diagonalizing (\ref{2b_hoping}) one obtains $e_g$ band dispersions with the total bandwidth $W=6t$.
\end{widetext}

The interaction term $H_{int}$  invariant over the cubic group symmetries   reads (see, e.g., Refs.~\onlinecite{Oles2000,Georges2013}):

\begin{align}\label{H_int_eg}
\hat{H}_{int}= &U\sum_{i,\alpha=a,b} \hat{n}_{i\alpha\uparrow}\hat{n}_{i\alpha\downarrow}+(U-2J_H)\sum_{i,a\neq b} \hat{n}_{ia\uparrow}\hat{n}_{ib\downarrow}  \nonumber  \\
 &+  (U-3J_H)\sum_{i\sigma} \hat{n}_{ia\sigma}\hat{n}_{ib\sigma}\\
 &+  J_H\sum_{i,a\neq b}(
f^{\dagger}_{ia\uparrow}f^{\dagger}_{ia\downarrow}f_{ib\downarrow}f_{ib\uparrow}-
f^{\dagger}_{ia\uparrow}f_{ia\downarrow}f^{\dagger}_{ib\downarrow}f_{ib\uparrow}), \nonumber
\end{align}
where $U$ and $J_H$ are the Coulomb and Hund's rule interactions, respectively.

We study the case of one-quarter filling, $Q=1$, for which  the model (\ref{Hab}) is relevant for a number of transition-metal compounds, for example, potassium copper fluorite KCuF$_3$ \cite{kugel_khomskii,Feiner1997,Oles2000} and rare-earth nickelates $R$NiO$_3$\cite{Balents2011,Park2012,Subedi2015}. Essentially the same model was studied  within DMFT in various parameter regimes to understand the behavior of nickelate-based heterostructures\cite{Hansmann2009,Han2011,Middey2016}.  The magnitude of super-exchange antiferromagnetic coupling is believed to be a crucial parameter controlling the physics of those heterostructures \cite{Hansmann2009,Hansmann2010}.

\begin{figure}[b]
	\centering
	\includegraphics[width=0.85\columnwidth]{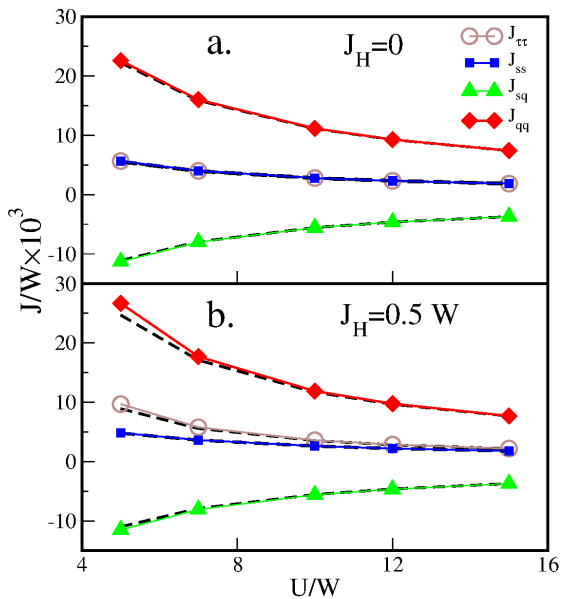}
	\caption{\label{fig:eg_model_J} Calculated superexchange inter-site interactions for the $e_g$ Hubbard model with  a. the Hund's rule coupling $J_H=0$; b. $J_H=0.5W$. The black dashed lines are the values obtained from analytical formulas (\ref{j_eg_an}).    }  
\end{figure}

We first carried out DMFT calculations employing the HIA with $J_H$ set either to 0 or to 0.5$W$ and $U$ being in the range from 5$W$ to 15$W$. The lower limit of $U$ is chosen to be above the critical value  $U_c=aW+3J$ (where the prefactor $a$ lies in the range from 1.5 to 2.5 depending on the lattice type under consideration) for the Mott transition in the two-band Hubbard model at quarter filing \cite{Ono2003,Rueff2014,Subedi2015}. Then the superexchange inter-site interactions at the first coordination shell between four one-electron states $|3z^2-r^2,\uparrow\rangle$, $|3z^2-r^2,\downarrow\rangle$,   $|x^2-y^2,\uparrow\rangle$, and $|x^2-y^2,\downarrow\rangle$ were computed in accordance with (\ref{V}). Finally, we employed eqs. (\ref{V_to_ts}) and (\ref{to_op}) to recast them into the standard Kugel-Khomskii\cite{kugel_khomskii} form of interacting spin-1/2 operators $\hat{\bf s}$ and $\hat{\tau}$ representing spin and orbital degrees of freedom, respectively ($\tau=1/2$ and $\tau=-1/2$ designate occupied $x^2-y^2$ and $3z^2-r^2$, respectively). The resulting effective Hamiltonian for the [001] bond, $\langle ij \rangle ||\hat{z}$, reads

\begin{align}\label{H_zz_2b}
\hat{H}_{eff}^{[001]}= & J_{ss}\sum_{\alpha}\hat{s}_{i\alpha}\hat{s}_{j\alpha}+
J_{\tau\tau}\hat{\tau}_{iz}\hat{\tau}_{jz}+
J_{qq}\sum_{\alpha}(\hat{s}_{i\alpha}\hat{\tau}_{iz})(\hat{s}_{j\alpha}\hat{\tau}_{jz}) \nonumber \\
& +J_{sq}\sum_{\alpha}
\left[\hat{s}_{i\alpha}(\hat{s}_{j\alpha}\hat{\tau}_{jz})+(\hat{s}_{i\alpha}\hat{\tau}_{iz})\hat{s}_{j\alpha}\right], 
\end{align}
where $J_{ss}$, $J_{\tau\tau}$, $J_{sq}$, and $J_{qq}$ are the spin-spin, orbital-orbital, spin-(spin-orbital) and (spin-orbital)-(spin-orbital) interactions defined in Sec.~\ref{sec_multipolar}, respectively, $\alpha$ runs over $x$, $y$, and $z$. As expected, the calculated effective Hamiltonians for  the [100] and [010] bonds are related by the cubic symmetry to $\hat{H}_{eff}^{[001]}$ and can be obtained from it  by the corresponding rotation in the $\hat{\tau}$ space, i. e., by substituting  
$\hat{\tau}_{z}$ in (\ref{H_zz_2b}) with $-\frac{1}{2}\hat{\tau}_z+\frac{\sqrt{3}}{2}\hat{\tau}_x$ and $-\frac{1}{2}\hat{\tau}_z-\frac{\sqrt{3}}{2}\hat{\tau}_x$, respectively.

The calculated values of $J_{ss}$, $J_{\tau\tau}$, $J_{sq}$, and $J_{qq}$ vs. $U$ are displayed in Fig.~\ref{fig:eg_model_J} together with the corresponding values of those superexchange interactions obtained from the analytical expressions  derived in Refs.~\cite{Feiner1997,Oles2000}:

\beq\label{j_eg_an}
 J_{ss}=J(1-\eta);\ J_{\tau\tau}=J(1+2\eta);\ J_{sq}=-J(2-\eta);\ J_{qq}=4J,
\eeq
where $J=\frac{t^2}{\tilde{U}}$, $\eta=\frac{2J_H}{\tilde{U}}$ with $\tilde{U}=U-J_H$ being the average Coulomb repulsion between $e_g$ electrons with opposite spins. One may note a perfect agreement between the calculated and analytical values in Fig.~\ref{fig:eg_model_J}a for the case $J_H=0$, for which eqs. (\ref{j_eg_an}) reduce to $J_{ss}=J_{\tau\tau}=- J_{sq}/2=J_{qq}/4=\frac{t^2}{U}$. In the case of $J_H=0.5W$ (Fig.~\ref{fig:eg_model_J}b) there are small discrepancies between the present approach and the analytical formulas (\ref{j_eg_an}) at low values of $U$. This is apparently due to the fact that the formulas (\ref{j_eg_an})  were derived\cite{Oles2000} by the first-order expansion in $\eta$ and become less accurate with increasing $J_H/U$.

\section{Spin and orbital ordering in KCrF$_3$}\label{KCrF3_sec}

In this section we calculate {\it ab initio} superexchange interactions for the cubic phase of KCrF$_3$ and then employ the resulting effective Hamiltonian to compute ordered phases and transition temperatures within the mean-field approximation.

First we carried out DFT+DMFT calculations of KCrF$_3$ using the linearized augmented plain-wave (LAPW) band structure method as implemented in the Wien2k\cite{Wien2k} code in conjunction with the DMFT and HIA implementations provided by the TRIQS library\cite{Parcollet2015,Aichhorn2016}. The Wannier orbitals representing correlated Cr 3$d$ states were constructed  using the projective approach of Ref.~\onlinecite{Aichhorn2009} from the Kohn-Sham (KS) states in the window [-2.7:2.7] eV around the Fermi level, this window encloses both $e_g$ and $t_{2g}$-like KS bands. The self-consistency over the charge density in the DFT+DMFT calculations was implemented as described in Ref.~\onlinecite{Aichhorn2011}, the spin-orbit coupling was neglected.

\begin{figure}
	\centering
	\includegraphics[width=0.85\columnwidth]{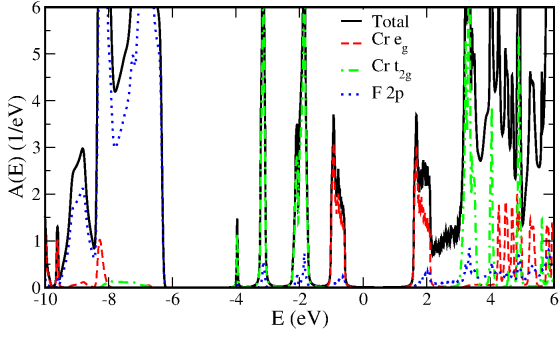}
	\caption{\label{fig:DOS} The total and projected spectral functions of KCrF$_3$ calculated by the DFT+DMFT method within the Hubbard-I approximation using $U=3.75$~eV. }  
\end{figure}

The rotationally-invariant local Coulomb repulsion between all five Cr 3$d$ orbitals was parametrized by the Slater integrals $F_0=U=3.75$~eV, as well as $F_2=6.44$~eV and $F_4=0.625 F_6=4.025$~eV corresponding to the Hund's rule coupling $J_H=0.75$~eV. Those values of  $F_0=U=3.75$ and $J_H=0.75$~eV were  computed for KCrF$_3$ in Ref.~\onlinecite{Autieri2014} using a constrained-LDA technique.  We also performed calculations with $U=5$~eV for the sake of comparison. We employed the fully-localized-limit form for the double counting correction term calculated with the nominal Cr 3$d$ shell occupancy of 4, this choice was shown to be appropriate for the HIA \cite{Pourovskii2007}.

KCrF$_3$ was calculated in its high-temperature cubic peroxide structure with the experimental\cite{Margadonna2007}  lattice parameter  of 4.23~\AA . We employed the atomic sphere radii of 2.5, 2.0 and 1.78 a.u. for K, Cr and F, respectively. The Brillouin zone (BZ) integration was carried out using 4000 {\bf k}-points in the full BZ, test calculations showed that increasing further the density of the {\bf k}-mesh had a negligible effect on the values of superexchange interactions.

Our DFT+DMFT calculations within HIA predict KCrF$_3$ to be a Mott insulator. Its spectral function features a Mott-Hubbard gap of about 2 eV, with the gap edges formed by $e_g$-like bands, see Fig.~\ref{fig:DOS}. The high-spin $t_{2g}^3e_g^1$ configuration with the total spin $S=2$ is predicted to be the ground-state multiplet of the Cr 3$d$ shell, as expected. Due to the orbital degeneracy of $3z^2-r^2$ and $x^2-y^2$ the total degeneracy of the ground-state multiplet is $2(2S+1)=10$.

Effective inter-site interactions (\ref{V}) between those 10 states belonging to the ground-state multiplet were then calculated in accordance with the approach of Secs.~\ref{local_fluct} and \ref{sec_grand_pot}. Then we again made use of eqs. (24) and (25) to recast them into the Kugel-Khomskii form. 

The calculated interactions between second nearest neighbors and beyond are at least two orders of magnitude smaller then those between the nearest neighbors and were neglected. The calculated superexchange Hamiltonian  between two nearest neighbors $i$ and $j$ along the [001] direction has the following form

\begin{widetext}
\begin{eqnarray}\label{H_zz_KCF3}
%\begin{tabular}{ll}
\hat{H}_{eff}^{[001]} & =J_{ss}\sum_{\alpha}\hat{s}_{i\alpha}\hat{s}_{j\alpha}+
J^{xy}_{\tau\tau}\sum_{\beta}\hat{\tau}_{i\beta}\hat{\tau}_{j\beta}+J_{\tau\tau}\hat{\tau}_{iz}\hat{\tau}_{jz}+
J_{sq}\sum_{\alpha}\left[\hat{s}_{i\alpha}(\hat{s}_{j\alpha}\hat{\tau}_{jz})+(\hat{s}_{i\alpha}\hat{\tau}_{iz})\hat{s}_{j\alpha}\right]+ \\ 
 & J^{xy}_{qq}\sum_{\alpha\beta}(\hat{s}_{i\alpha}\hat{\tau}_{i\beta})
(\hat{s}_{j\alpha}\hat{\tau}_{j\beta})+
J_{qq}\sum_{\alpha}(\hat{s}_{i\alpha}\hat{\tau}_{iz})(\hat{s}_{j\alpha}\hat{\tau}_{jz}), \nonumber
%\end{tabular}
\end{eqnarray}
\end{widetext}
where  $\alpha$ and $\beta$ run over $x$, $y$, $z$  and $x$, $y$, respectively. The spin operators $\hat{s}_{i\alpha}$ act in the $S=2$ space of the total spin of the site $i$,  the $\tau=1/2$ and $\tau=-1/2$ quantum numbers designate the  $t_{2g}^3[x^2-y^2]$ and $t_{2g}^3[3z^2-r^2]$ shell configurations, respectively. The meaning of $J_{ss}$, $J_{\tau\tau}$, $J_{sq}$ and $J_{qq}$ is the same as in eq. (\ref{H_zz_2b}) of Sec.~\ref{hub_2b}. Comparing (\ref{H_zz_KCF3}) to  (\ref{H_zz_2b}) one notices the appearance of new $J^{xy}_{\tau\tau}$ and $J^{xy}_{qq}$ terms. Because there is no inter-orbital hopping within the $e_g$ subshell along the $z$ axis those terms should be related to virtual hopping of $t_{2g}$ electrons (hence, they are absent from the pure $e_g$ model, eq.~\ref{H_zz_2b} )\footnote{While there is no symmetry reason for the interactions of $\hat{\tau}_{ix}\hat{\tau}_{jx}$ and $\hat{\tau}_{iy}\hat{\tau}_{jy}$ to be equal we found them to be almost coinciding.}. 

\begin{table}
\centering
\caption{\label{table:J} Calculated Cr-Cr nearest-neighbor interactions along the [001] direction, in meV .}
    \begin{tabular}{c| c c c c c c }
U (eV)        & $J_{ss}$ & $J_{\tau\tau}$ & $J^{xy}_{\tau\tau}$ & $J_{sq}$ & $J_{qq}$ &  $J^{xy}_{qq}$ \\ \hline
3.75   & 0.94 & 37.3 & 1.71 & -1.77 & 7.12 & 0.28 \\
5 & 0.96 & 24.7 & 1.20 & -1.43 & 4.93 & 0.21 \\
   \end{tabular}
\end{table}

In Table~\ref{table:J} we list the values of the inter-site interactions calculated with $U=3.75$~eV and $U=5$~eV. One sees that the orbital-orbital  $J_{\tau\tau}$ and (spin-orbital)-(spin-orbital) $J_{qq}$ interactions are the most significant ones, even if one takes into account the different lengths of $\tau=1/2$ and $S=2$ spins. These interactions exhibit a strong reduction upon increasing $U$ and decreasing $J_H/U$, cf. (\ref{j_eg_an}). The  (spin-orbital)-(spin) term $J_{sq}$ is also significant. $J^{xy}_{\tau\tau}$ and $J^{xy}_{qq}$ are more than one order of magnitude smaller than  $J_{\tau\tau}$ and $J_{qq}$, respectively.

We have then solved the calculated nearest-neighbor superexchange Hamiltonian defined by  eq. \ref{H_zz_KCF3} (the nearest-neighbor interactions along the [100] and [010] directions are obtained from (\ref{H_zz_KCF3}) using rotations in the $\tau$ space as described in Sec.~\ref{hub_2b}) within the mean-field approximation using McPhase package \cite{Rotter2004} obtaining the total and free energies as well as stable ordered phases as a function of temperature.

\begin{figure}[b]
	\centering
	\includegraphics[width=0.85\columnwidth]{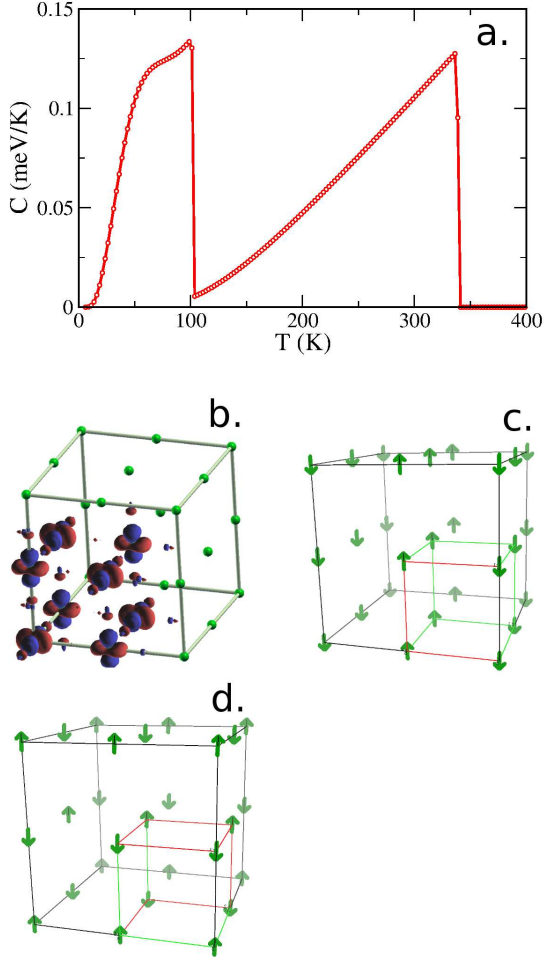}
	\caption{\label{fig:C_and_struct} a. Specific heat (per formula unit) as a function of temperature obtained by solving the Hamiltionian (\ref{H_zz_KCF3}) with the values of superexchange interactions calculated at $U=3.75$~eV. b. The G-type antiferro-orbital order obtained below $T_{OO}=340$~K  (plotted by XCrysDen\cite{Kokalj2003}, the real-space representation of the orbitals is generated with the help of the wplot\cite{Kunes2010} program). c. The A-type anti-ferromagnetic phase, stable below  $T_{N}=102$~K, obtained with the interactions calculated with $U=3.75$~eV d.  The C-type anti-ferromagnetic phase obtained using the interactions calculated with $U=5$~eV.}  
\end{figure}

The calculated temperature dependence of the  specific heat (see Fig.~\ref{fig:C_and_struct}a)   features two clear phase transitions at temperatures of 340 and 102~K. The high-temperature one is an orbital-ordering transition, the obtained  antiferro-orbital structure is displayed in Fig.~\ref{fig:C_and_struct}b. The occupied $e_g$ states on two inequivalent sites, (which are the nearest neighbors in the simple-cubic Cr sublattice, see Fig.~\ref{fig:C_and_struct}b) in this structure can be written as 

\begin{align}\label{wf_orbital}
|\theta\rangle=\cos \theta |3z^2-r^2\rangle + \sin \theta |x^2-y^2\rangle, \\
\label{wf_orbital_1}
|\theta_1\rangle=-\sin \theta_1 |3z^2-r^2\rangle + \cos \theta_1 |x^2-y^2\rangle,
\end{align}
with $\theta=\theta_1$, hence, the obtained structure corresponds to a G-type antiferro-orbital order with the empty $e_g$ orbital on the site one being occupied on the site two \footnote{The notation used in eqs. (\ref{wf_orbital}) and (\ref{wf_orbital_1}) follows that used in Ref.~\onlinecite{Margadonna2007}. Another representation of the G-type antiferro-orbital order is more standard in the case of LaMnO$_3$ (see, for examle, Ref.~\onlinecite{Feinberg1998}) and obtained by substituting  $\theta$ in (\ref{wf_orbital}) and $\theta_1$ in  (\ref{wf_orbital_1}) with $\theta/2$ and $-\theta_1/2$, respectively}. The actual value of the angle $\theta$ is not defined by the Hamiltonian (\ref{H_zz_KCF3}) in the absence of the spin ordering, experimentally it is  fixed by the tetragonal lattice distortion and equal to 30$^{\circ}$\cite{Margadonna2007}. In fact, neglecting the lattice distortion leads to a strongly underestimated value of the temperature $T_{OO}$ for the orbital ordering compared to experimental 973$~K$. The same result was obtained by Autieri {\it et al.} \cite{Autieri2014}  using direct DMFT+QMC calculations and was shown to be due to the on-site crystal field splitting, the renormalization of hopping integrals due to the tetragonal (and subsequent monoclinic) distortion had an insignificant effect on $T_{OO}$.

The low-temperature transition at $T_{N}=102$~K is due to ordering of  Cr spins into the AFM A-type structure shown in Fig.~\ref{fig:C_and_struct}c. This structure consists of an antiferromagnetic stacking of ferromagnetically-ordered $xz$ layers, with each Cr site having four in-plain neighbours with the same spin and two out-of-plain ones with the opposite spin. This, in fact, is the collinear spin structure observed experimentally in KCrF$_3$. The obtained N\'eel temperature is in  good agreement with experimental value of 80~K\cite{Xiao2010}, if one takes into account the usual mean-field overestimation of ordering temperatures. Hence, one sees that once the orbital order sets in the superexchange is able to account for the value of $T_N$ and observed collinear magnetic structure even without including lattice distortions \footnote{A possibility for slight non-collinearity as the one present in KCrF$_3$ between 46 and 80~K was not considered in the mean-field solution of the effective Hamiltonian. Also our simulations neglect the spin-orbit couping and, hence, are not able to reproduce the spin canting observed below 9~K}.

We have also performed the same mean-field calculations with the effective interactions computed with $U=5.0$~eV, obtaining the same orbitally-ordered structure at somewhat lower temperature of 225~K. The obtained low-temperature spin structure is, however, different, it is of the C-type and consists of an anitferromagnetic stacking of ferromagnetically-ordered [101] plains. Hence, each site has two nearest-neighbors with the same spin and four with the opposite one, see Fig.~\ref{fig:C_and_struct}d.

In order to clarify the origin of this change of magnetic order with increasing $U$ one may carry out a simple estimate of the energy of the A-type and C-type AFM spin structures. First, keeping in (\ref{H_zz_KCF3}) only the most important $J_{ss}$ and $J_{qq}$ contributions and summing over all nearest-neighbors one obtains for the mean-field spin-ordering energy (per formula unit, f.u.) of the A-type structure

\begin{widetext}
\beq\label{E_SO}
\frac{E_{SO}^{A-type}}{S^2}  = J_{ss}+J_{qq}\left(\langle \hat{\tau}_{iz} \rangle \langle \hat{\tau}_{jz} \rangle - \frac{\sqrt{3}}{2}\left(\langle \hat{\tau}_{iz} \rangle \langle \hat{\tau}_{jx} \rangle + \langle \hat{\tau}_{ix} \rangle \langle \hat{\tau}_{jz} \rangle \right) \right),
\eeq
where $S^2$ is the overall spin factor, which for the case of high-spin  Cr$^{2+}$ can be rather well approximated by the square of its classical length, $S^2=4$ at the full saturation, $i$ and $j$ label two sublattices of the  G-type antiferro-orbital structure. The energy of the C-type structure $E_{SO}^{C-type}$ is given by the same expression with the minus sign. 
\end{widetext} 

The energies of the   ferromagnetic (FM) and G-type AFM (all nearest neighbors having the opposite spin) phases are 
\beq\label{E_FM_AFM}
\pm 3S^2\left(J_{ss}+\frac{J_{qq}}{2}( \langle \hat{\tau}_{iz} \rangle \langle \hat{\tau}_{jz} \rangle + \langle \hat{\tau}_{ix} \rangle \langle \hat{\tau}_{jx} \rangle)\right),
\eeq
 where the plus/minus sign is for the FM/AFM case, respectively. One may notice that $ \langle \hat{\tau}_{iz} \rangle \langle \hat{\tau}_{jz} \rangle + \langle \hat{\tau}_{ix} \rangle \langle \hat{\tau}_{jx} \rangle$ is always equal to -1/4  for the fully saturated G-type antiferro-orbital order and does  not dependent on the angle $\theta$ in eqs. (\ref{wf_orbital}) and (\ref{wf_orbital_1}). Hence, the total energy of the FM and G-type AFM order is also independent of $\theta$.

 Assuming a fully-saturated G-type antiferro-orbital order, i. e. $ \langle \hat{\tau}_{ix} \rangle = \pm \sqrt{1/4- \langle \hat{\tau}_{iz} \rangle^2}$, $\langle \hat{\tau}_{jx} \rangle = -\langle \hat{\tau}_{ix} \rangle$ and  $\langle \hat{\tau}_{jz} \rangle = - \langle \hat{\tau}_{iz} \rangle$, and minimizing (\ref{E_SO}) one obtains  $E_{SO}^{A-type}=4J_{ss}-3J_{qq}/2$ with the orbital state fixed at $ \langle \hat{\tau}_{iz} \rangle=\sqrt{3}/4$ and  $ \langle \hat{\tau}_{ix} \rangle=-1/4$, defined by $\theta=15^{\circ}$ in (\ref{wf_orbital}). For the C-type structure one has  $E_{SO}^{C-type}=-4J_{ss}-J_{qq}/2$ and the orbital state locked at $\langle \hat{\tau}_{iz} \rangle=1/4$ and  $ \langle \hat{\tau}_{ix} \rangle=\sqrt{3}/4$, corresponding to $\theta=30^{\circ}$ (which is, in fact, the experimental orbital state in tetragonal KCrF$_3$). The energies of the FM and  G-type AFM  orders do not dependent on $\theta$ as explained above and are equal to $\pm 3(4J_{ss}-J_{qq}/2)$, respectively.   Hence, one sees that the spin order is defined by the ratio of $J_{qq}/J_{ss}$, which increases with decreasing $U$ (increasing $J_H/U$), see Table~\ref{table:J}. For $U=5$~eV and $U=3.75$~eV one obtains for the energy difference $E_{SO}^{A-type}-E_{SO}^{C-type}$ the values of 2.75 and 0.4~meV per f.u., respectively. Hence, at the realistic value of $U=3.75$~eV those two structures are almost degenerate, though $E_{SO}^{C-type}$ is still the most stable. The G-type AFM and FM structures are always higher in energy, in particular, for $U=3.75$~eV their energies are 6.7 and 7.9 meV per f.u. above $E_{SO}^{C-type}$.

\begin{figure}[b]
\centering
\includegraphics[width=0.85\columnwidth]{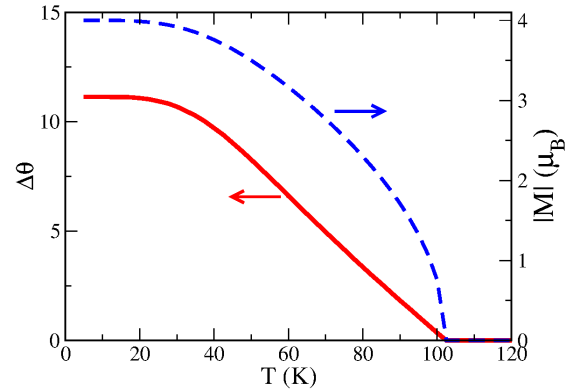}
\caption{\label{fig:theta} The spin magnetic moment and orbital misalignment angle $\Delta \theta$ as a function of temperature in the A-type structure obtained for $U=3.75$~eV.}  
\end{figure}

Further analysis shows that the A-type structure is stabilized at  $U=3.75$~eV due to the $J_{sq}$ term, which upon the onset of antiferromagnitism acts as a canting field in the orbital space. For the C-type structure it takes the form $|J_{qs}|S^2\left[\langle\hat{\tau}_{iz}\rangle + \langle\hat{\tau}_{jz}\rangle +\sqrt{3}(\langle\hat{\tau}_{ix}\rangle + \langle\hat{\tau}_{jx}\rangle)\right]$ and one may show that under the corresponding G-type antiferro-orbital order given by $\theta=30^{\circ}$ it is not active as long as $J_{qq}S^2+8J_{sq}S^2+2J_{\tau\tau} > 0$. The system stays in the same antiferro-orbital state with $\theta=30^{\circ}$, for which the contribution of the $J_{sq}$ term to the energy is zero. For the A-type structure it takes the same form with the minus sign, but now under the different orbital state given by $ \langle \hat{\tau}_{iz} \rangle=\sqrt{3}/4$ and  $ \langle \hat{\tau}_{ix} \rangle=-1/4$  ($\theta=15^{\circ}$) it does play a role leading to a loss of the perfect G-type antiferro-orbital order. Namely, upon the onset of the A-type spin order the angles $\theta$ and $\theta_1$ defining the corresponding orbital states (\ref{wf_orbital}) and (\ref{wf_orbital_1}) on two sublattices start deviating from each other, the corresponding loss in the orbital ordering energy is compensated by the "orbital field" due to $J_{sq}$. The corresponding difference $\Delta \theta = \theta_1 - \theta$ (extracted from  mean-field solution of the full effective Hamiltonian, eq.~\ref{H_zz_KCF3}) grows with decreasing temperature due to increasing spin moment, as shown in Fig.~\ref{fig:theta}. This orbital misalignment stops increasing once the magnetic moment fully saturates below approximately 30~K. The total gain in energy due to this misalignment of about -1.4 meV/(f.u.) is rather small compared to the total energy of the spin-orbital ordering of -22.9 meV/(f.u.), but it is sufficient to stabilize the A-type antiferromagnetic order.

Previously the interplay of orbital ordering and the A-type AFM structure has been intensively studied in the case of the Mn peroxide LaMnO$_3$, where Jahn-Teller lattice distortions were proposed to be at the origin of this AFM structure, see, for example, Refs.~\onlinecite{Solovyev1996,Sawada1997,Feinberg1998}. The dependence of the relative stability of different magnetic phases of LaMnO$_3$ on $J_H/U$ was previously demonstrated in a model study of Ref.~\cite{Maezono1998}.  KCrF$_3$ features some similarities to this  system, though in LaMnO$_3$ the orbital order is of the C-type instead of the G-type. Direct {\it ab initio} DFT+U calculations\cite{Giovannetti2008,Xing2014} for cubic KCrF$_3$ predicted a ferro-orbital order to be stabilized in conjunction with the A-type AFM, in disagreement with our results and experiment. Apparently, this is due to an incorrect relative scale of spin- and orbital-ordering energies in DFT+U within the local spin-density approximation, where the orbital order is seen to be induced by the underline AFM state\cite{Giovannetti2008}. As one sees from Table~\ref{table:J}, the inter-orbital superexchange is the strongest interaction, hence the AFM state emerges well below $T_{OO}$ from an almost completely saturated G-type antiferro-orbital order, in agreement with experiment. In our description the angle $\theta$, which defines the orbital state (\ref{wf_orbital}) , is fixed by the lowest-energy AFM order, in real KCrF$_3$ it is rather fixed by the distorted lattice. The tetragonal distortion favors the C-type AFM in accordance with our calculations, however, experimentally the magnetic order emerges in the lower-temperature monoclinic structure, in which the orbital state is possibly more favorable to the A-type magnetic order. It is interesting to observe  that  even in the absence of any lattice distortions the feedback effect described above leads to a canted orbital order in conjunction with the A-type AFM. Experimentally one may also expect to observe an additional small titling/distortion of the CrF$_6$ octahedra upon the onset of antiferromagnitism, though our calculations predict a rather small energy scale associated with this process, of the order of 1 meV.

\section{Summary}

We have presented a method for computing inter-site exchange interactions in correlated materials in the framework of the DFT+DMFT in conjunction with the Hubbard-I approximation to the DMFT self-energy. The expressions for inter-site interactions are derived by considering the first-order change in the DFT+DMFT grand potential to simultaneous small fluctuations on two atomic sites with respect to their symmetry-unbroken paramagnetic configuration.  The resulting expression (\ref{V}) combines the variational derivatives of the Hubbard-I self-energy (\ref{Self_fluct}) over a given fluctuation in the on-site density matrix with the DMFT inter-site Green's functions (\ref{G_off}). The method is benchmarked by applying it to the well-known cases of one-band and two-band $e_g$ Hubbard models on the simple-cubic 3$d$ lattice.

The presented technique has been already employed to compute spin-spin superexchange interactions in cubic and quasi-two-dimensional tetragonal  TM oxides \cite{Horvat2015}. Here we have applied it to a more complex case of spin-orbital ordering in KCrF$_3$ in its parent undistorted peroxide structure. We obtained an effective Hamiltonian (\ref{H_zz_KCF3})  featuring strong antiferro-orbital nearest-neighbor interactions and a complex anisotropic coupling between orbital and spin moments. By solving it within the mean-field approximation we found  the onset of a G-type orbital order at a significantly lower temperature as compared to experiment. In contrast, the appearence of experimentally-observed A-type antiferromagnetic structure is predicted at  $T_N=102$~K in good agreement with experiment. The onset of A-type antiferromagnetism is explained by purely superexchange mechanism as arising due to  an interplay of the spin-spin and (spin-orbit)-(spin-orbit) inter-site couplings in conjunction with a canting of the G-type antiferro-orbital order. Further applications of this technique to the tetragonal and monoclinic structures of KCrF$_3$ should help to clarify whether this mechanism for the stabilization of the A-type magnetic structure is qualitatively affected by the lattice distortions. 

The present method is promising for applications to a wide range of strongly-correlated materials, like spin-orbital order in TM oxides and florides as well as multipolar ordering due to localized $f$ shells in rare-earth and actinide materials. It would be interesting to consider its generalizations beyond the Hubbard-I approximation to widen its range of applicability to materials located close to the Mott point like, e.g.,  rare-earth nickelates. One might also try to extend the present formalism in order to incorporate contributions to spin-orbital ordering from Jahn-Teller-type distortions.

\section{Acknowlegments}
The author is grateful to A. Georges for his invaluable help in the beginning of this work. J. Mravlje and O. Peil are acknowledged for useful discussions.  The author acknowledges the financial support of
the Ministry of Education and Science of the Russian Federation in the framework of
Increase Competitiveness Program of NUST MISiS (No. K3-2015-038) as well as computational resources provided by   the National Supercomputer Centre in Link\"oping (NSC)  at Swedish National Infrastructure
for Computing (SNIC).

%\bibliography{mag_int}

\begin{thebibliography}{69}
	\expandafter\ifx\csname natexlab\endcsname\relax\def\natexlab#1{#1}\fi
	\expandafter\ifx\csname bibnamefont\endcsname\relax
	\def\bibnamefont#1{#1}\fi
	\expandafter\ifx\csname bibfnamefont\endcsname\relax
	\def\bibfnamefont#1{#1}\fi
	\expandafter\ifx\csname citenamefont\endcsname\relax
	\def\citenamefont#1{#1}\fi
	\expandafter\ifx\csname url\endcsname\relax
	\def\url#1{\texttt{#1}}\fi
	\expandafter\ifx\csname urlprefix\endcsname\relax\def\urlprefix{URL }\fi
	\providecommand{\bibinfo}[2]{#2}
	\providecommand{\eprint}[2][]{\url{#2}}
	
	\bibitem[{\citenamefont{K~I~Kugel' and Khomskii}(1982)}]{kugel_khomskii}
	\bibinfo{author}{\bibfnamefont{K.~I.} \bibnamefont{K~I~Kugel'}}
	\bibnamefont{and} \bibinfo{author}{\bibfnamefont{D.~I.}
		\bibnamefont{Khomskii}}, \bibinfo{journal}{Sov. Phys. Uspekhi}
	\textbf{\bibinfo{volume}{25}}, \bibinfo{pages}{231} (\bibinfo{year}{1982}).
	
	\bibitem[{\citenamefont{Shiina et~al.}(1997)\citenamefont{Shiina, Shiba, and
			Thalmeier}}]{Shiina1997}
	\bibinfo{author}{\bibfnamefont{R.}~\bibnamefont{Shiina}},
	\bibinfo{author}{\bibfnamefont{H.}~\bibnamefont{Shiba}}, \bibnamefont{and}
	\bibinfo{author}{\bibfnamefont{P.}~\bibnamefont{Thalmeier}},
	\bibinfo{journal}{Journal of the Physical Society of Japan}
	\textbf{\bibinfo{volume}{66}}, \bibinfo{pages}{1741} (\bibinfo{year}{1997}).
	
	\bibitem[{\citenamefont{Kuramoto et~al.}(2009)\citenamefont{Kuramoto, Kusunose,
			and Kiss}}]{Kuramoto2009}
	\bibinfo{author}{\bibfnamefont{Y.}~\bibnamefont{Kuramoto}},
	\bibinfo{author}{\bibfnamefont{H.}~\bibnamefont{Kusunose}}, \bibnamefont{and}
	\bibinfo{author}{\bibfnamefont{A.}~\bibnamefont{Kiss}},
	\bibinfo{journal}{Journal of the Physical Society of Japan}
	\textbf{\bibinfo{volume}{78}}, \bibinfo{pages}{072001}
	(\bibinfo{year}{2009}).
	
	\bibitem[{\citenamefont{Santini et~al.}(2009)\citenamefont{Santini, Carretta,
			Amoretti, Caciuffo, Magnani, and Lander}}]{Santini2009}
	\bibinfo{author}{\bibfnamefont{P.}~\bibnamefont{Santini}},
	\bibinfo{author}{\bibfnamefont{S.}~\bibnamefont{Carretta}},
	\bibinfo{author}{\bibfnamefont{G.}~\bibnamefont{Amoretti}},
	\bibinfo{author}{\bibfnamefont{R.}~\bibnamefont{Caciuffo}},
	\bibinfo{author}{\bibfnamefont{N.}~\bibnamefont{Magnani}}, \bibnamefont{and}
	\bibinfo{author}{\bibfnamefont{G.~H.} \bibnamefont{Lander}},
	\bibinfo{journal}{Rev. Mod. Phys.} \textbf{\bibinfo{volume}{81}},
	\bibinfo{pages}{807} (\bibinfo{year}{2009}).
	
	\bibitem[{\citenamefont{Mydosh and Oppeneer}(2011)}]{Mydosh2011}
	\bibinfo{author}{\bibfnamefont{J.~A.} \bibnamefont{Mydosh}} \bibnamefont{and}
	\bibinfo{author}{\bibfnamefont{P.~M.} \bibnamefont{Oppeneer}},
	\bibinfo{journal}{Rev. Mod. Phys.} \textbf{\bibinfo{volume}{83}},
	\bibinfo{pages}{1301} (\bibinfo{year}{2011}).
	
	\bibitem[{\citenamefont{Anisimov et~al.}(1997)\citenamefont{Anisimov,
			Poteryaev, Korotin, Anokhin, and Kotliar}}]{anisimov97}
	\bibinfo{author}{\bibfnamefont{V.~I.} \bibnamefont{Anisimov}},
	\bibinfo{author}{\bibfnamefont{A.~I.} \bibnamefont{Poteryaev}},
	\bibinfo{author}{\bibfnamefont{M.~A.} \bibnamefont{Korotin}},
	\bibinfo{author}{\bibfnamefont{A.~O.} \bibnamefont{Anokhin}},
	\bibnamefont{and} \bibinfo{author}{\bibfnamefont{G.}~\bibnamefont{Kotliar}},
	\bibinfo{journal}{J. Phys.: Condens. Matter} \textbf{\bibinfo{volume}{9}},
	\bibinfo{pages}{7359} (\bibinfo{year}{1997}).
	
	\bibitem[{\citenamefont{Kotliar et~al.}(2006)\citenamefont{Kotliar, Savrasov,
			Haule, Oudovenko, Parcollet, and Marianetti}}]{Kotliar2006}
	\bibinfo{author}{\bibfnamefont{G.}~\bibnamefont{Kotliar}},
	\bibinfo{author}{\bibfnamefont{S.~Y.} \bibnamefont{Savrasov}},
	\bibinfo{author}{\bibfnamefont{K.}~\bibnamefont{Haule}},
	\bibinfo{author}{\bibfnamefont{V.~S.} \bibnamefont{Oudovenko}},
	\bibinfo{author}{\bibfnamefont{O.}~\bibnamefont{Parcollet}},
	\bibnamefont{and} \bibinfo{author}{\bibfnamefont{C.~A.}
		\bibnamefont{Marianetti}}, \bibinfo{journal}{Rev. Mod. Phys.}
	\textbf{\bibinfo{volume}{78}}, \bibinfo{pages}{865} (\bibinfo{year}{2006}).
	
	\bibitem[{\citenamefont{Georges et~al.}(1996)\citenamefont{Georges, Kotliar,
			Krauth, and Rozenberg}}]{Georges1996}
	\bibinfo{author}{\bibfnamefont{A.}~\bibnamefont{Georges}},
	\bibinfo{author}{\bibfnamefont{G.}~\bibnamefont{Kotliar}},
	\bibinfo{author}{\bibfnamefont{W.}~\bibnamefont{Krauth}}, \bibnamefont{and}
	\bibinfo{author}{\bibfnamefont{M.~J.} \bibnamefont{Rozenberg}},
	\bibinfo{journal}{Rev. Mod. Phys.} \textbf{\bibinfo{volume}{68}},
	\bibinfo{pages}{13} (\bibinfo{year}{1996}).
	
	\bibitem[{\citenamefont{Rohringer et~al.}(2011)\citenamefont{Rohringer, Toschi,
			Katanin, and Held}}]{Rohringer2011}
	\bibinfo{author}{\bibfnamefont{G.}~\bibnamefont{Rohringer}},
	\bibinfo{author}{\bibfnamefont{A.}~\bibnamefont{Toschi}},
	\bibinfo{author}{\bibfnamefont{A.}~\bibnamefont{Katanin}}, \bibnamefont{and}
	\bibinfo{author}{\bibfnamefont{K.}~\bibnamefont{Held}},
	\bibinfo{journal}{Phys. Rev. Lett.} \textbf{\bibinfo{volume}{107}},
	\bibinfo{pages}{256402} (\bibinfo{year}{2011}).
	
	\bibitem[{\citenamefont{Hirschmeier et~al.}(2015)\citenamefont{Hirschmeier,
			Hafermann, Gull, Lichtenstein, and Antipov}}]{Hirschmeier2015}
	\bibinfo{author}{\bibfnamefont{D.}~\bibnamefont{Hirschmeier}},
	\bibinfo{author}{\bibfnamefont{H.}~\bibnamefont{Hafermann}},
	\bibinfo{author}{\bibfnamefont{E.}~\bibnamefont{Gull}},
	\bibinfo{author}{\bibfnamefont{A.~I.} \bibnamefont{Lichtenstein}},
	\bibnamefont{and} \bibinfo{author}{\bibfnamefont{A.~E.}
		\bibnamefont{Antipov}}, \bibinfo{journal}{Phys. Rev. B}
	\textbf{\bibinfo{volume}{92}}, \bibinfo{pages}{144409}
	(\bibinfo{year}{2015}).
	
	\bibitem[{\citenamefont{Maier et~al.}(2005)\citenamefont{Maier, Jarrell,
			Schulthess, Kent, and White}}]{Maier2005}
	\bibinfo{author}{\bibfnamefont{T.~A.} \bibnamefont{Maier}},
	\bibinfo{author}{\bibfnamefont{M.}~\bibnamefont{Jarrell}},
	\bibinfo{author}{\bibfnamefont{T.~C.} \bibnamefont{Schulthess}},
	\bibinfo{author}{\bibfnamefont{P.~R.~C.} \bibnamefont{Kent}},
	\bibnamefont{and} \bibinfo{author}{\bibfnamefont{J.~B.} \bibnamefont{White}},
	\bibinfo{journal}{Phys. Rev. Lett.} \textbf{\bibinfo{volume}{95}},
	\bibinfo{pages}{237001} (\bibinfo{year}{2005}).
	
	\bibitem[{\citenamefont{Sch\"afer et~al.}(2015)\citenamefont{Sch\"afer, Geles,
			Rost, Rohringer, Arrigoni, Held, Bl\"umer, Aichhorn, and
			Toschi}}]{Schafer2015}
	\bibinfo{author}{\bibfnamefont{T.}~\bibnamefont{Sch\"afer}},
	\bibinfo{author}{\bibfnamefont{F.}~\bibnamefont{Geles}},
	\bibinfo{author}{\bibfnamefont{D.}~\bibnamefont{Rost}},
	\bibinfo{author}{\bibfnamefont{G.}~\bibnamefont{Rohringer}},
	\bibinfo{author}{\bibfnamefont{E.}~\bibnamefont{Arrigoni}},
	\bibinfo{author}{\bibfnamefont{K.}~\bibnamefont{Held}},
	\bibinfo{author}{\bibfnamefont{N.}~\bibnamefont{Bl\"umer}},
	\bibinfo{author}{\bibfnamefont{M.}~\bibnamefont{Aichhorn}}, \bibnamefont{and}
	\bibinfo{author}{\bibfnamefont{A.}~\bibnamefont{Toschi}},
	\bibinfo{journal}{Phys. Rev. B} \textbf{\bibinfo{volume}{91}},
	\bibinfo{pages}{125109} (\bibinfo{year}{2015}).
	
	\bibitem[{\citenamefont{Ayral and Parcollet}(2015)}]{Ayral2015}
	\bibinfo{author}{\bibfnamefont{T.}~\bibnamefont{Ayral}} \bibnamefont{and}
	\bibinfo{author}{\bibfnamefont{O.}~\bibnamefont{Parcollet}},
	\bibinfo{journal}{Phys. Rev. B} \textbf{\bibinfo{volume}{92}},
	\bibinfo{pages}{115109} (\bibinfo{year}{2015}).
	
	\bibitem[{\citenamefont{Horvat et~al.}()\citenamefont{Horvat, Pourovskii,
			Aichhorn, and Mravlje}}]{Horvat2015}
	\bibinfo{author}{\bibfnamefont{A.}~\bibnamefont{Horvat}},
	\bibinfo{author}{\bibfnamefont{L.}~\bibnamefont{Pourovskii}},
	\bibinfo{author}{\bibfnamefont{M.}~\bibnamefont{Aichhorn}}, \bibnamefont{and}
	\bibinfo{author}{\bibfnamefont{J.}~\bibnamefont{Mravlje}},
	\bibinfo{note}{arXiv:1501.03033 (unpublished)}.
	
	\bibitem[{\citenamefont{Prange and Korenman}(1979)}]{Prange1979}
	\bibinfo{author}{\bibfnamefont{R.~E.} \bibnamefont{Prange}} \bibnamefont{and}
	\bibinfo{author}{\bibfnamefont{V.}~\bibnamefont{Korenman}},
	\bibinfo{journal}{Phys. Rev. B} \textbf{\bibinfo{volume}{19}},
	\bibinfo{pages}{4691} (\bibinfo{year}{1979}).
	
	\bibitem[{\citenamefont{Wang et~al.}(1982)\citenamefont{Wang, Prange, and
			Korenman}}]{Wang1982}
	\bibinfo{author}{\bibfnamefont{C.~S.} \bibnamefont{Wang}},
	\bibinfo{author}{\bibfnamefont{R.~E.} \bibnamefont{Prange}},
	\bibnamefont{and} \bibinfo{author}{\bibfnamefont{V.}~\bibnamefont{Korenman}},
	\bibinfo{journal}{Phys. Rev. B} \textbf{\bibinfo{volume}{25}},
	\bibinfo{pages}{5766} (\bibinfo{year}{1982}).
	
	\bibitem[{\citenamefont{Oguchi et~al.}(1983{\natexlab{a}})\citenamefont{Oguchi,
			Terakura, and Hamada}}]{Oguchi1983}
	\bibinfo{author}{\bibfnamefont{T.}~\bibnamefont{Oguchi}},
	\bibinfo{author}{\bibfnamefont{K.}~\bibnamefont{Terakura}}, \bibnamefont{and}
	\bibinfo{author}{\bibfnamefont{N.}~\bibnamefont{Hamada}},
	\bibinfo{journal}{Journal of Physics F: Metal Physics}
	\textbf{\bibinfo{volume}{13}}, \bibinfo{pages}{145}
	(\bibinfo{year}{1983}{\natexlab{a}}).
	
	\bibitem[{\citenamefont{Oguchi et~al.}(1983{\natexlab{b}})\citenamefont{Oguchi,
			Terakura, and Williams}}]{Oguchi1983prb}
	\bibinfo{author}{\bibfnamefont{T.}~\bibnamefont{Oguchi}},
	\bibinfo{author}{\bibfnamefont{K.}~\bibnamefont{Terakura}}, \bibnamefont{and}
	\bibinfo{author}{\bibfnamefont{A.~R.} \bibnamefont{Williams}},
	\bibinfo{journal}{Phys. Rev. B} \textbf{\bibinfo{volume}{28}},
	\bibinfo{pages}{6443} (\bibinfo{year}{1983}{\natexlab{b}}).
	
	\bibitem[{\citenamefont{Liechtenstein et~al.}(1987)\citenamefont{Liechtenstein,
			Katsnelson, Antropov, and Gubanov}}]{Liechtenstein1987}
	\bibinfo{author}{\bibfnamefont{A.}~\bibnamefont{Liechtenstein}},
	\bibinfo{author}{\bibfnamefont{M.}~\bibnamefont{Katsnelson}},
	\bibinfo{author}{\bibfnamefont{V.}~\bibnamefont{Antropov}}, \bibnamefont{and}
	\bibinfo{author}{\bibfnamefont{V.}~\bibnamefont{Gubanov}},
	\bibinfo{journal}{Journal of Magnetism and Magnetic Materials}
	\textbf{\bibinfo{volume}{67}}, \bibinfo{pages}{65 } (\bibinfo{year}{1987}).
	
	\bibitem[{\citenamefont{Bruno}(2003)}]{Bruno2003}
	\bibinfo{author}{\bibfnamefont{P.}~\bibnamefont{Bruno}},
	\bibinfo{journal}{Phys. Rev. Lett.} \textbf{\bibinfo{volume}{90}},
	\bibinfo{pages}{087205} (\bibinfo{year}{2003}).
	
	\bibitem[{\citenamefont{Ruban et~al.}(2004)\citenamefont{Ruban, Shallcross,
			Simak, and Skriver}}]{Ruban2004}
	\bibinfo{author}{\bibfnamefont{A.~V.} \bibnamefont{Ruban}},
	\bibinfo{author}{\bibfnamefont{S.}~\bibnamefont{Shallcross}},
	\bibinfo{author}{\bibfnamefont{S.~I.} \bibnamefont{Simak}}, \bibnamefont{and}
	\bibinfo{author}{\bibfnamefont{H.~L.} \bibnamefont{Skriver}},
	\bibinfo{journal}{Phys. Rev. B} \textbf{\bibinfo{volume}{70}},
	\bibinfo{pages}{125115} (\bibinfo{year}{2004}).
	
	\bibitem[{\citenamefont{Solovyev et~al.}(1995)\citenamefont{Solovyev,
			Dederichs, and Mertig}}]{Solovyev1995}
	\bibinfo{author}{\bibfnamefont{I.~V.} \bibnamefont{Solovyev}},
	\bibinfo{author}{\bibfnamefont{P.~H.} \bibnamefont{Dederichs}},
	\bibnamefont{and} \bibinfo{author}{\bibfnamefont{I.}~\bibnamefont{Mertig}},
	\bibinfo{journal}{Phys. Rev. B} \textbf{\bibinfo{volume}{52}},
	\bibinfo{pages}{13419} (\bibinfo{year}{1995}).
	
	\bibitem[{\citenamefont{Solovyev et~al.}(1996)\citenamefont{Solovyev, Hamada,
			and Terakura}}]{Solovyev1996}
	\bibinfo{author}{\bibfnamefont{I.}~\bibnamefont{Solovyev}},
	\bibinfo{author}{\bibfnamefont{N.}~\bibnamefont{Hamada}}, \bibnamefont{and}
	\bibinfo{author}{\bibfnamefont{K.}~\bibnamefont{Terakura}},
	\bibinfo{journal}{Phys. Rev. Lett.} \textbf{\bibinfo{volume}{76}},
	\bibinfo{pages}{4825} (\bibinfo{year}{1996}).
	
	\bibitem[{\citenamefont{Katsnelson and Lichtenstein}(2000)}]{Katsnelson2000}
	\bibinfo{author}{\bibfnamefont{M.~I.} \bibnamefont{Katsnelson}}
	\bibnamefont{and} \bibinfo{author}{\bibfnamefont{A.~I.}
		\bibnamefont{Lichtenstein}}, \bibinfo{journal}{Phys. Rev. B}
	\textbf{\bibinfo{volume}{61}}, \bibinfo{pages}{8906} (\bibinfo{year}{2000}).
	
	\bibitem[{\citenamefont{Pi et~al.}(2014)\citenamefont{Pi, Nanguneri, and
			Savrasov}}]{Pi2014}
	\bibinfo{author}{\bibfnamefont{S.-T.} \bibnamefont{Pi}},
	\bibinfo{author}{\bibfnamefont{R.}~\bibnamefont{Nanguneri}},
	\bibnamefont{and} \bibinfo{author}{\bibfnamefont{S.}~\bibnamefont{Savrasov}},
	\bibinfo{journal}{Phys. Rev. Lett.} \textbf{\bibinfo{volume}{112}},
	\bibinfo{pages}{077203} (\bibinfo{year}{2014}).
	
	\bibitem[{\citenamefont{Secchi et~al.}(2015)\citenamefont{Secchi, Lichtenstein,
			and Katsnelson}}]{Secchi2015}
	\bibinfo{author}{\bibfnamefont{A.}~\bibnamefont{Secchi}},
	\bibinfo{author}{\bibfnamefont{A.}~\bibnamefont{Lichtenstein}},
	\bibnamefont{and}
	\bibinfo{author}{\bibfnamefont{M.}~\bibnamefont{Katsnelson}},
	\bibinfo{journal}{Annals of Physics} \textbf{\bibinfo{volume}{360}},
	\bibinfo{pages}{61 } (\bibinfo{year}{2015}).
	
	\bibitem[{\citenamefont{Hubbard}(1963)}]{hubbard_1}
	\bibinfo{author}{\bibfnamefont{J.}~\bibnamefont{Hubbard}},
	\bibinfo{journal}{Proc R. Soc. Lond. A} \textbf{\bibinfo{volume}{276}},
	\bibinfo{pages}{238} (\bibinfo{year}{1963}).
	
	\bibitem[{\citenamefont{Margadonna and Karotsis}(2007)}]{Margadonna2007}
	\bibinfo{author}{\bibfnamefont{S.}~\bibnamefont{Margadonna}} \bibnamefont{and}
	\bibinfo{author}{\bibfnamefont{G.}~\bibnamefont{Karotsis}},
	\bibinfo{journal}{J. Mater. Chem.} \textbf{\bibinfo{volume}{17}},
	\bibinfo{pages}{2013} (\bibinfo{year}{2007}).
	
	\bibitem[{\citenamefont{Margadonna and Karotsis}(2006)}]{Margadonna2006}
	\bibinfo{author}{\bibfnamefont{S.}~\bibnamefont{Margadonna}} \bibnamefont{and}
	\bibinfo{author}{\bibfnamefont{G.}~\bibnamefont{Karotsis}},
	\bibinfo{journal}{Journal of the American Chemical Society}
	\textbf{\bibinfo{volume}{128}}, \bibinfo{pages}{16436}
	(\bibinfo{year}{2006}).
	
	\bibitem[{\citenamefont{Xiao et~al.}(2010)\citenamefont{Xiao, Su, Li, Kumar,
			Mittal, Persson, Senyshyn, Gross, and Brueckel}}]{Xiao2010}
	\bibinfo{author}{\bibfnamefont{Y.}~\bibnamefont{Xiao}},
	\bibinfo{author}{\bibfnamefont{Y.}~\bibnamefont{Su}},
	\bibinfo{author}{\bibfnamefont{H.-F.} \bibnamefont{Li}},
	\bibinfo{author}{\bibfnamefont{C.~M.~N.} \bibnamefont{Kumar}},
	\bibinfo{author}{\bibfnamefont{R.}~\bibnamefont{Mittal}},
	\bibinfo{author}{\bibfnamefont{J.}~\bibnamefont{Persson}},
	\bibinfo{author}{\bibfnamefont{A.}~\bibnamefont{Senyshyn}},
	\bibinfo{author}{\bibfnamefont{K.}~\bibnamefont{Gross}}, \bibnamefont{and}
	\bibinfo{author}{\bibfnamefont{T.}~\bibnamefont{Brueckel}},
	\bibinfo{journal}{Phys. Rev. B} \textbf{\bibinfo{volume}{82}},
	\bibinfo{pages}{094437} (\bibinfo{year}{2010}).
	
	\bibitem[{\citenamefont{Autieri et~al.}(2014)\citenamefont{Autieri, Koch, and
			Pavarini}}]{Autieri2014}
	\bibinfo{author}{\bibfnamefont{C.}~\bibnamefont{Autieri}},
	\bibinfo{author}{\bibfnamefont{E.}~\bibnamefont{Koch}}, \bibnamefont{and}
	\bibinfo{author}{\bibfnamefont{E.}~\bibnamefont{Pavarini}},
	\bibinfo{journal}{Phys. Rev. B} \textbf{\bibinfo{volume}{89}},
	\bibinfo{pages}{155109} (\bibinfo{year}{2014}).
	
	\bibitem[{\citenamefont{Giovannetti et~al.}(2008)\citenamefont{Giovannetti,
			Margadonna, and van~den Brink}}]{Giovannetti2008}
	\bibinfo{author}{\bibfnamefont{G.}~\bibnamefont{Giovannetti}},
	\bibinfo{author}{\bibfnamefont{S.}~\bibnamefont{Margadonna}},
	\bibnamefont{and} \bibinfo{author}{\bibfnamefont{J.}~\bibnamefont{van~den
			Brink}}, \bibinfo{journal}{Phys. Rev. B} \textbf{\bibinfo{volume}{77}},
	\bibinfo{pages}{075113} (\bibinfo{year}{2008}).
	
	\bibitem[{\citenamefont{Xing et~al.}(2014)\citenamefont{Xing, Liang-Bin,
			Huo-Xi, Fei, Chun-Zhong, and Gang}}]{Xing2014}
	\bibinfo{author}{\bibfnamefont{M.}~\bibnamefont{Xing}},
	\bibinfo{author}{\bibfnamefont{X.}~\bibnamefont{Liang-Bin}},
	\bibinfo{author}{\bibfnamefont{X.}~\bibnamefont{Huo-Xi}},
	\bibinfo{author}{\bibfnamefont{D.}~\bibnamefont{Fei}},
	\bibinfo{author}{\bibfnamefont{W.}~\bibnamefont{Chun-Zhong}},
	\bibnamefont{and} \bibinfo{author}{\bibfnamefont{C.}~\bibnamefont{Gang}},
	\bibinfo{journal}{Chinese Physics B} \textbf{\bibinfo{volume}{23}},
	\bibinfo{eid}{37401} (\bibinfo{year}{2014}).
	
	\bibitem[{\citenamefont{{Pourovskii} et~al.}(2007)\citenamefont{{Pourovskii},
			{Amadon}, {Biermann}, and {Georges}}}]{Pourovskii2007}
	\bibinfo{author}{\bibfnamefont{L.~V.} \bibnamefont{{Pourovskii}}},
	\bibinfo{author}{\bibfnamefont{B.}~\bibnamefont{{Amadon}}},
	\bibinfo{author}{\bibfnamefont{S.}~\bibnamefont{{Biermann}}},
	\bibnamefont{and}
	\bibinfo{author}{\bibfnamefont{A.}~\bibnamefont{{Georges}}},
	\bibinfo{journal}{Phys. Rev. B} \textbf{\bibinfo{volume}{76}},
	\bibinfo{pages}{235101} (\bibinfo{year}{2007}).
	
	\bibitem[{\citenamefont{Savrasov and Kotliar}(2004)}]{Savrasov2004}
	\bibinfo{author}{\bibfnamefont{S.~Y.} \bibnamefont{Savrasov}} \bibnamefont{and}
	\bibinfo{author}{\bibfnamefont{G.}~\bibnamefont{Kotliar}},
	\bibinfo{journal}{Phys. Rev. B} \textbf{\bibinfo{volume}{69}},
	\bibinfo{pages}{245101} (\bibinfo{year}{2004}).
	
	\bibitem[{\citenamefont{Georges}(2004)}]{Georges2004}
	\bibinfo{author}{\bibfnamefont{A.}~\bibnamefont{Georges}},
	\bibinfo{journal}{AIP Conference Proceedings} \textbf{\bibinfo{volume}{715}},
	\bibinfo{pages}{3} (\bibinfo{year}{2004}).
	
	\bibitem[{\citenamefont{Mackintosh and Andersen}(1980)}]{force_theorem}
	\bibinfo{author}{\bibfnamefont{A.~R.} \bibnamefont{Mackintosh}}
	\bibnamefont{and} \bibinfo{author}{\bibfnamefont{O.~K.}
		\bibnamefont{Andersen}}, in \emph{\bibinfo{booktitle}{Electrons at the Fermi
			Surface}}, edited by
	\bibinfo{editor}{\bibfnamefont{M.}~\bibnamefont{Springford}}
	(\bibinfo{publisher}{Cambridge University Press},
	\bibinfo{address}{Cambridge, England}, \bibinfo{year}{1980}), p.
	\bibinfo{pages}{145}.
	
	\bibitem[{\citenamefont{Solovyev and Terakura}(1998)}]{Solovyev1998}
	\bibinfo{author}{\bibfnamefont{I.~V.} \bibnamefont{Solovyev}} \bibnamefont{and}
	\bibinfo{author}{\bibfnamefont{K.}~\bibnamefont{Terakura}},
	\bibinfo{journal}{Phys. Rev. B} \textbf{\bibinfo{volume}{58}},
	\bibinfo{pages}{15496} (\bibinfo{year}{1998}).
	
	\bibitem[{\citenamefont{Plefka}(1982)}]{Plefka1982}
	\bibinfo{author}{\bibfnamefont{T.}~\bibnamefont{Plefka}},
	\bibinfo{journal}{Journal of Physics A: Mathematical and General}
	\textbf{\bibinfo{volume}{15}}, \bibinfo{pages}{1971} (\bibinfo{year}{1982}).
	
	\bibitem[{\citenamefont{Georges and Yedidia}(1991)}]{Georges1991}
	\bibinfo{author}{\bibfnamefont{A.}~\bibnamefont{Georges}} \bibnamefont{and}
	\bibinfo{author}{\bibfnamefont{J.~S.} \bibnamefont{Yedidia}},
	\bibinfo{journal}{Journal of Physics A: Mathematical and General}
	\textbf{\bibinfo{volume}{24}}, \bibinfo{pages}{2173} (\bibinfo{year}{1991}).
	
	\bibitem[{\citenamefont{Blum}(1996)}]{Blum_DM}
	\bibinfo{author}{\bibfnamefont{K.}~\bibnamefont{Blum}},
	\emph{\bibinfo{title}{Density matrix theory and applications}}
	(\bibinfo{publisher}{Plenum Press, New York}, \bibinfo{year}{1996}).
	
	\bibitem[{\citenamefont{Parcollet et~al.}(2015)\citenamefont{Parcollet,
			Ferrero, Ayral, Hafermann, Krivenko, Messio, and Seth}}]{Parcollet2015}
	\bibinfo{author}{\bibfnamefont{O.}~\bibnamefont{Parcollet}},
	\bibinfo{author}{\bibfnamefont{M.}~\bibnamefont{Ferrero}},
	\bibinfo{author}{\bibfnamefont{T.}~\bibnamefont{Ayral}},
	\bibinfo{author}{\bibfnamefont{H.}~\bibnamefont{Hafermann}},
	\bibinfo{author}{\bibfnamefont{I.}~\bibnamefont{Krivenko}},
	\bibinfo{author}{\bibfnamefont{L.}~\bibnamefont{Messio}}, \bibnamefont{and}
	\bibinfo{author}{\bibfnamefont{P.}~\bibnamefont{Seth}},
	\bibinfo{journal}{Computer Physics Communications}
	\textbf{\bibinfo{volume}{196}}, \bibinfo{pages}{398 } (\bibinfo{year}{2015}),
	\urlprefix\url{http://ipht.cea.fr/triqs/}.
	
	\bibitem[{\citenamefont{Ulmke et~al.}(1995)\citenamefont{Ulmke,
			Jani\ifmmode~\check{s}\else \v{s}\fi{}, and Vollhardt}}]{Ulmke1995}
	\bibinfo{author}{\bibfnamefont{M.}~\bibnamefont{Ulmke}},
	\bibinfo{author}{\bibfnamefont{V.}~\bibnamefont{Jani\ifmmode~\check{s}\else
			\v{s}\fi{}}}, \bibnamefont{and}
	\bibinfo{author}{\bibfnamefont{D.}~\bibnamefont{Vollhardt}},
	\bibinfo{journal}{Phys. Rev. B} \textbf{\bibinfo{volume}{51}},
	\bibinfo{pages}{10411} (\bibinfo{year}{1995}).
	
	\bibitem[{\citenamefont{Jarrell}(1992)}]{Jarell1992}
	\bibinfo{author}{\bibfnamefont{M.}~\bibnamefont{Jarrell}},
	\bibinfo{journal}{Phys. Rev. Lett.} \textbf{\bibinfo{volume}{69}},
	\bibinfo{pages}{168} (\bibinfo{year}{1992}).
	
	\bibitem[{\citenamefont{Strack and Vollhardt}(1992)}]{Rainer1992}
	\bibinfo{author}{\bibfnamefont{R.}~\bibnamefont{Strack}} \bibnamefont{and}
	\bibinfo{author}{\bibfnamefont{D.}~\bibnamefont{Vollhardt}},
	\bibinfo{journal}{Phys. Rev. B} \textbf{\bibinfo{volume}{46}},
	\bibinfo{pages}{13852} (\bibinfo{year}{1992}).
	
	\bibitem[{\citenamefont{Sangiovanni et~al.}(2006)\citenamefont{Sangiovanni,
			Toschi, Koch, Held, Capone, Castellani, Gunnarsson, Mo, Allen, Kim
			et~al.}}]{Sangiovanni2006}
	\bibinfo{author}{\bibfnamefont{G.}~\bibnamefont{Sangiovanni}},
	\bibinfo{author}{\bibfnamefont{A.}~\bibnamefont{Toschi}},
	\bibinfo{author}{\bibfnamefont{E.}~\bibnamefont{Koch}},
	\bibinfo{author}{\bibfnamefont{K.}~\bibnamefont{Held}},
	\bibinfo{author}{\bibfnamefont{M.}~\bibnamefont{Capone}},
	\bibinfo{author}{\bibfnamefont{C.}~\bibnamefont{Castellani}},
	\bibinfo{author}{\bibfnamefont{O.}~\bibnamefont{Gunnarsson}},
	\bibinfo{author}{\bibfnamefont{S.-K.} \bibnamefont{Mo}},
	\bibinfo{author}{\bibfnamefont{J.~W.} \bibnamefont{Allen}},
	\bibinfo{author}{\bibfnamefont{H.-D.} \bibnamefont{Kim}},
	\bibnamefont{et~al.}, \bibinfo{journal}{Phys. Rev. B}
	\textbf{\bibinfo{volume}{73}}, \bibinfo{pages}{205121}
	(\bibinfo{year}{2006}).
	
	\bibitem[{\citenamefont{Rozenberg et~al.}(1994)\citenamefont{Rozenberg,
			Kotliar, and Zhang}}]{Rozenberg1994}
	\bibinfo{author}{\bibfnamefont{M.~J.} \bibnamefont{Rozenberg}},
	\bibinfo{author}{\bibfnamefont{G.}~\bibnamefont{Kotliar}}, \bibnamefont{and}
	\bibinfo{author}{\bibfnamefont{X.~Y.} \bibnamefont{Zhang}},
	\bibinfo{journal}{Phys. Rev. B} \textbf{\bibinfo{volume}{49}},
	\bibinfo{pages}{10181} (\bibinfo{year}{1994}).
	
	\bibitem[{\citenamefont{Ole\ifmmode~\acute{s}\else \'{s}\fi{}
			et~al.}(2000)\citenamefont{Ole\ifmmode~\acute{s}\else \'{s}\fi{}, Feiner, and
			Zaanen}}]{Oles2000}
	\bibinfo{author}{\bibfnamefont{A.~M.} \bibnamefont{Ole\ifmmode~\acute{s}\else
			\'{s}\fi{}}}, \bibinfo{author}{\bibfnamefont{L.~F.} \bibnamefont{Feiner}},
	\bibnamefont{and} \bibinfo{author}{\bibfnamefont{J.}~\bibnamefont{Zaanen}},
	\bibinfo{journal}{Phys. Rev. B} \textbf{\bibinfo{volume}{61}},
	\bibinfo{pages}{6257} (\bibinfo{year}{2000}).
	
	\bibitem[{\citenamefont{Georges et~al.}(2013)\citenamefont{Georges, de' Medici,
			and Mravlje}}]{Georges2013}
	\bibinfo{author}{\bibfnamefont{A.}~\bibnamefont{Georges}},
	\bibinfo{author}{\bibfnamefont{L.}~\bibnamefont{de' Medici}},
	\bibnamefont{and} \bibinfo{author}{\bibfnamefont{J.}~\bibnamefont{Mravlje}},
	\bibinfo{journal}{Annual Review of Condensed Matter Physics}
	\textbf{\bibinfo{volume}{4}}, \bibinfo{pages}{137} (\bibinfo{year}{2013}).
	
	\bibitem[{\citenamefont{Feiner et~al.}(1997)\citenamefont{Feiner,
			Ole\ifmmode~\acute{s}\else \'{s}\fi{}, and Zaanen}}]{Feiner1997}
	\bibinfo{author}{\bibfnamefont{L.~F.} \bibnamefont{Feiner}},
	\bibinfo{author}{\bibfnamefont{A.~M.} \bibnamefont{Ole\ifmmode~\acute{s}\else
			\'{s}\fi{}}}, \bibnamefont{and}
	\bibinfo{author}{\bibfnamefont{J.}~\bibnamefont{Zaanen}},
	\bibinfo{journal}{Phys. Rev. Lett.} \textbf{\bibinfo{volume}{78}},
	\bibinfo{pages}{2799} (\bibinfo{year}{1997}).
	
	\bibitem[{\citenamefont{Lee et~al.}(2011)\citenamefont{Lee, Chen, and
			Balents}}]{Balents2011}
	\bibinfo{author}{\bibfnamefont{S.~B.} \bibnamefont{Lee}},
	\bibinfo{author}{\bibfnamefont{R.}~\bibnamefont{Chen}}, \bibnamefont{and}
	\bibinfo{author}{\bibfnamefont{L.}~\bibnamefont{Balents}},
	\bibinfo{journal}{Phys. Rev. B} \textbf{\bibinfo{volume}{84}},
	\bibinfo{pages}{165119} (\bibinfo{year}{2011}).
	
	\bibitem[{\citenamefont{Park et~al.}(2012)\citenamefont{Park, Millis, and
			Marianetti}}]{Park2012}
	\bibinfo{author}{\bibfnamefont{H.}~\bibnamefont{Park}},
	\bibinfo{author}{\bibfnamefont{A.~J.} \bibnamefont{Millis}},
	\bibnamefont{and} \bibinfo{author}{\bibfnamefont{C.~A.}
		\bibnamefont{Marianetti}}, \bibinfo{journal}{Phys. Rev. Lett.}
	\textbf{\bibinfo{volume}{109}}, \bibinfo{pages}{156402}
	(\bibinfo{year}{2012}).
	
	\bibitem[{\citenamefont{Subedi et~al.}(2015)\citenamefont{Subedi, Peil, and
			Georges}}]{Subedi2015}
	\bibinfo{author}{\bibfnamefont{A.}~\bibnamefont{Subedi}},
	\bibinfo{author}{\bibfnamefont{O.~E.} \bibnamefont{Peil}}, \bibnamefont{and}
	\bibinfo{author}{\bibfnamefont{A.}~\bibnamefont{Georges}},
	\bibinfo{journal}{Phys. Rev. B} \textbf{\bibinfo{volume}{91}},
	\bibinfo{pages}{075128} (\bibinfo{year}{2015}).
	
	\bibitem[{\citenamefont{Hansmann et~al.}(2009)\citenamefont{Hansmann, Yang,
			Toschi, Khaliullin, Andersen, and Held}}]{Hansmann2009}
	\bibinfo{author}{\bibfnamefont{P.}~\bibnamefont{Hansmann}},
	\bibinfo{author}{\bibfnamefont{X.}~\bibnamefont{Yang}},
	\bibinfo{author}{\bibfnamefont{A.}~\bibnamefont{Toschi}},
	\bibinfo{author}{\bibfnamefont{G.}~\bibnamefont{Khaliullin}},
	\bibinfo{author}{\bibfnamefont{O.~K.} \bibnamefont{Andersen}},
	\bibnamefont{and} \bibinfo{author}{\bibfnamefont{K.}~\bibnamefont{Held}},
	\bibinfo{journal}{Phys. Rev. Lett.} \textbf{\bibinfo{volume}{103}},
	\bibinfo{pages}{016401} (\bibinfo{year}{2009}).
	
	\bibitem[{\citenamefont{Han et~al.}(2011)\citenamefont{Han, Wang, Marianetti,
			and Millis}}]{Han2011}
	\bibinfo{author}{\bibfnamefont{M.~J.} \bibnamefont{Han}},
	\bibinfo{author}{\bibfnamefont{X.}~\bibnamefont{Wang}},
	\bibinfo{author}{\bibfnamefont{C.~A.} \bibnamefont{Marianetti}},
	\bibnamefont{and} \bibinfo{author}{\bibfnamefont{A.~J.}
		\bibnamefont{Millis}}, \bibinfo{journal}{Phys. Rev. Lett.}
	\textbf{\bibinfo{volume}{107}}, \bibinfo{pages}{206804}
	(\bibinfo{year}{2011}).
	
	\bibitem[{\citenamefont{Middey et~al.}(2016)\citenamefont{Middey, Chakhalian,
			Mahadevan, Freeland, Millis, and Sarma}}]{Middey2016}
	\bibinfo{author}{\bibfnamefont{S.}~\bibnamefont{Middey}},
	\bibinfo{author}{\bibfnamefont{J.}~\bibnamefont{Chakhalian}},
	\bibinfo{author}{\bibfnamefont{P.}~\bibnamefont{Mahadevan}},
	\bibinfo{author}{\bibfnamefont{J.}~\bibnamefont{Freeland}},
	\bibinfo{author}{\bibfnamefont{A.}~\bibnamefont{Millis}}, \bibnamefont{and}
	\bibinfo{author}{\bibfnamefont{D.}~\bibnamefont{Sarma}},
	\bibinfo{journal}{Annual Review of Materials Research}
	\textbf{\bibinfo{volume}{46}}, \bibinfo{pages}{305} (\bibinfo{year}{2016}).
	
	\bibitem[{\citenamefont{Hansmann et~al.}(2010)\citenamefont{Hansmann, Toschi,
			Yang, Andersen, and Held}}]{Hansmann2010}
	\bibinfo{author}{\bibfnamefont{P.}~\bibnamefont{Hansmann}},
	\bibinfo{author}{\bibfnamefont{A.}~\bibnamefont{Toschi}},
	\bibinfo{author}{\bibfnamefont{X.}~\bibnamefont{Yang}},
	\bibinfo{author}{\bibfnamefont{O.~K.} \bibnamefont{Andersen}},
	\bibnamefont{and} \bibinfo{author}{\bibfnamefont{K.}~\bibnamefont{Held}},
	\bibinfo{journal}{Phys. Rev. B} \textbf{\bibinfo{volume}{82}},
	\bibinfo{pages}{235123} (\bibinfo{year}{2010}).
	
	\bibitem[{\citenamefont{\ifmmode~\bar{O}\else \={O}\fi{}no
			et~al.}(2003)\citenamefont{\ifmmode~\bar{O}\else \={O}\fi{}no, Potthoff, and
			Bulla}}]{Ono2003}
	\bibinfo{author}{\bibfnamefont{Y.}~\bibnamefont{\ifmmode~\bar{O}\else
			\={O}\fi{}no}}, \bibinfo{author}{\bibfnamefont{M.}~\bibnamefont{Potthoff}},
	\bibnamefont{and} \bibinfo{author}{\bibfnamefont{R.}~\bibnamefont{Bulla}},
	\bibinfo{journal}{Phys. Rev. B} \textbf{\bibinfo{volume}{67}},
	\bibinfo{pages}{035119} (\bibinfo{year}{2003}).
	
	\bibitem[{\citenamefont{R\"uegg et~al.}(2014)\citenamefont{R\"uegg, Hung, Gull,
			and Fiete}}]{Rueff2014}
	\bibinfo{author}{\bibfnamefont{A.}~\bibnamefont{R\"uegg}},
	\bibinfo{author}{\bibfnamefont{H.-H.} \bibnamefont{Hung}},
	\bibinfo{author}{\bibfnamefont{E.}~\bibnamefont{Gull}}, \bibnamefont{and}
	\bibinfo{author}{\bibfnamefont{G.~A.} \bibnamefont{Fiete}},
	\bibinfo{journal}{Phys. Rev. B} \textbf{\bibinfo{volume}{89}},
	\bibinfo{pages}{085122} (\bibinfo{year}{2014}).
	
	\bibitem[{\citenamefont{Blaha et~al.}(2001)\citenamefont{Blaha, Schwarz,
			Madsen, Kvasnicka, and Luitz}}]{Wien2k}
	\bibinfo{author}{\bibfnamefont{P.}~\bibnamefont{Blaha}},
	\bibinfo{author}{\bibfnamefont{K.}~\bibnamefont{Schwarz}},
	\bibinfo{author}{\bibfnamefont{G.}~\bibnamefont{Madsen}},
	\bibinfo{author}{\bibfnamefont{D.}~\bibnamefont{Kvasnicka}},
	\bibnamefont{and} \bibinfo{author}{\bibfnamefont{J.}~\bibnamefont{Luitz}},
	\emph{\bibinfo{title}{WIEN2k, An augmented Plane Wave + Local Orbitals
			Program for Calculating Crystal Properties}} (\bibinfo{publisher}{Techn.
		Universitat Wien, Austria, ISBN 3-9501031-1-2.}, \bibinfo{year}{2001}).
	
	\bibitem[{\citenamefont{Aichhorn et~al.}(2016)\citenamefont{Aichhorn,
			Pourovskii, Seth, Vildosola, Zingl, Peil, Deng, Mravlje, Kraberger, Martins
			et~al.}}]{Aichhorn2016}
	\bibinfo{author}{\bibfnamefont{M.}~\bibnamefont{Aichhorn}},
	\bibinfo{author}{\bibfnamefont{L.}~\bibnamefont{Pourovskii}},
	\bibinfo{author}{\bibfnamefont{P.}~\bibnamefont{Seth}},
	\bibinfo{author}{\bibfnamefont{V.}~\bibnamefont{Vildosola}},
	\bibinfo{author}{\bibfnamefont{M.}~\bibnamefont{Zingl}},
	\bibinfo{author}{\bibfnamefont{O.~E.} \bibnamefont{Peil}},
	\bibinfo{author}{\bibfnamefont{X.}~\bibnamefont{Deng}},
	\bibinfo{author}{\bibfnamefont{J.}~\bibnamefont{Mravlje}},
	\bibinfo{author}{\bibfnamefont{G.~J.} \bibnamefont{Kraberger}},
	\bibinfo{author}{\bibfnamefont{C.}~\bibnamefont{Martins}},
	\bibnamefont{et~al.}, \bibinfo{journal}{Computer Physics Communications}
	\textbf{\bibinfo{volume}{204}}, \bibinfo{pages}{200 } (\bibinfo{year}{2016}).
	
	\bibitem[{\citenamefont{Aichhorn et~al.}(2009)\citenamefont{Aichhorn,
			Pourovskii, Vildosola, Ferrero, Parcollet, Miyake, Georges, and
			Biermann}}]{Aichhorn2009}
	\bibinfo{author}{\bibfnamefont{M.}~\bibnamefont{Aichhorn}},
	\bibinfo{author}{\bibfnamefont{L.}~\bibnamefont{Pourovskii}},
	\bibinfo{author}{\bibfnamefont{V.}~\bibnamefont{Vildosola}},
	\bibinfo{author}{\bibfnamefont{M.}~\bibnamefont{Ferrero}},
	\bibinfo{author}{\bibfnamefont{O.}~\bibnamefont{Parcollet}},
	\bibinfo{author}{\bibfnamefont{T.}~\bibnamefont{Miyake}},
	\bibinfo{author}{\bibfnamefont{A.}~\bibnamefont{Georges}}, \bibnamefont{and}
	\bibinfo{author}{\bibfnamefont{S.}~\bibnamefont{Biermann}},
	\bibinfo{journal}{Phys. Rev. B} \textbf{\bibinfo{volume}{80}},
	\bibinfo{pages}{085101} (\bibinfo{year}{2009}).
	
	\bibitem[{\citenamefont{Aichhorn et~al.}(2011)\citenamefont{Aichhorn,
			Pourovskii, and Georges}}]{Aichhorn2011}
	\bibinfo{author}{\bibfnamefont{M.}~\bibnamefont{Aichhorn}},
	\bibinfo{author}{\bibfnamefont{L.}~\bibnamefont{Pourovskii}},
	\bibnamefont{and} \bibinfo{author}{\bibfnamefont{A.}~\bibnamefont{Georges}},
	\bibinfo{journal}{Phys. Rev. B} \textbf{\bibinfo{volume}{84}},
	\bibinfo{pages}{054529} (\bibinfo{year}{2011}).
	
	\bibitem[{\citenamefont{Rotter}(2004)}]{Rotter2004}
	\bibinfo{author}{\bibfnamefont{M.}~\bibnamefont{Rotter}},
	\bibinfo{journal}{Journal of Magnetism and Magnetic Materials}
	\textbf{\bibinfo{volume}{272-276, Supplement}}, \bibinfo{pages}{E481 }
	(\bibinfo{year}{2004}), \urlprefix\url{http://www.mcphase.de/}.
	
	\bibitem[{\citenamefont{Kokalj}(2003)}]{Kokalj2003}
	\bibinfo{author}{\bibfnamefont{A.}~\bibnamefont{Kokalj}},
	\bibinfo{journal}{Computational Materials Science}
	\textbf{\bibinfo{volume}{28}}, \bibinfo{pages}{155 } (\bibinfo{year}{2003}).
	
	\bibitem[{\citenamefont{Kune\v{s} et~al.}(2010)\citenamefont{Kune\v{s}, Arita,
			Wissgott, Toschi, Ikeda, and Held}}]{Kunes2010}
	\bibinfo{author}{\bibfnamefont{J.}~\bibnamefont{Kune\v{s}}},
	\bibinfo{author}{\bibfnamefont{R.}~\bibnamefont{Arita}},
	\bibinfo{author}{\bibfnamefont{P.}~\bibnamefont{Wissgott}},
	\bibinfo{author}{\bibfnamefont{A.}~\bibnamefont{Toschi}},
	\bibinfo{author}{\bibfnamefont{H.}~\bibnamefont{Ikeda}}, \bibnamefont{and}
	\bibinfo{author}{\bibfnamefont{K.}~\bibnamefont{Held}},
	\bibinfo{journal}{Computer Physics Communications}
	\textbf{\bibinfo{volume}{181}}, \bibinfo{pages}{1888 }
	(\bibinfo{year}{2010}).
	
	\bibitem[{\citenamefont{Sawada et~al.}(1997)\citenamefont{Sawada, Morikawa,
			Terakura, and Hamada}}]{Sawada1997}
	\bibinfo{author}{\bibfnamefont{H.}~\bibnamefont{Sawada}},
	\bibinfo{author}{\bibfnamefont{Y.}~\bibnamefont{Morikawa}},
	\bibinfo{author}{\bibfnamefont{K.}~\bibnamefont{Terakura}}, \bibnamefont{and}
	\bibinfo{author}{\bibfnamefont{N.}~\bibnamefont{Hamada}},
	\bibinfo{journal}{Phys. Rev. B} \textbf{\bibinfo{volume}{56}},
	\bibinfo{pages}{12154} (\bibinfo{year}{1997}).
	
	\bibitem[{\citenamefont{Feinberg et~al.}(1998)\citenamefont{Feinberg, Germain,
			Grilli, and Seibold}}]{Feinberg1998}
	\bibinfo{author}{\bibfnamefont{D.}~\bibnamefont{Feinberg}},
	\bibinfo{author}{\bibfnamefont{P.}~\bibnamefont{Germain}},
	\bibinfo{author}{\bibfnamefont{M.}~\bibnamefont{Grilli}}, \bibnamefont{and}
	\bibinfo{author}{\bibfnamefont{G.}~\bibnamefont{Seibold}},
	\bibinfo{journal}{Phys. Rev. B} \textbf{\bibinfo{volume}{57}},
	\bibinfo{pages}{R5583} (\bibinfo{year}{1998}).
	
	\bibitem[{\citenamefont{Maezono et~al.}(1998)\citenamefont{Maezono, Ishihara,
			and Nagaosa}}]{Maezono1998}
	\bibinfo{author}{\bibfnamefont{R.}~\bibnamefont{Maezono}},
	\bibinfo{author}{\bibfnamefont{S.}~\bibnamefont{Ishihara}}, \bibnamefont{and}
	\bibinfo{author}{\bibfnamefont{N.}~\bibnamefont{Nagaosa}},
	\bibinfo{journal}{Phys. Rev. B} \textbf{\bibinfo{volume}{58}},
	\bibinfo{pages}{11583} (\bibinfo{year}{1998}).
	
\end{thebibliography}

\end{document}